\documentclass[aoas]{imsart}

\RequirePackage{amsthm,amsmath,amsfonts,amssymb}
\RequirePackage[authoryear]{natbib}
\RequirePackage[colorlinks,citecolor=blue,urlcolor=blue]{hyperref}
\RequirePackage{graphicx}

\startlocaldefs
\usepackage{tikz}
\usetikzlibrary{positioning,fit,backgrounds,arrows.meta,calc}
\tikzset{
  block/.style={draw, thick, rectangle, rounded corners, fill=white, inner sep=6pt},
  arrow/.style={-Latex, thick}
}
\usepackage{xcolor}
\usepackage{placeins}
\usepackage[T1]{fontenc}
\usepackage{type1cm}   
\usepackage{algorithm}
\usepackage{algpseudocode}
\usepackage{float}

\usepackage{booktabs}
\usepackage{wrapfig}
\usepackage{capt-of}
\endlocaldefs

\begin{document}

\begin{frontmatter}

\title{A TWO-LEVEL PLACKETT-LUCE MODEL FOR PREFERENCE MODELING IN SMART MOBILITY PLATFORMS}
\thankstext{T1}{Preprint version.}

\begin{aug}
\author[A]{\fnms{M. SANTOS-PASCUAL}~\snm{}\ead[label=e1]{miguel.santos@icmat.es}},
\author[B]{\fnms{D. RÍOS INSUA}~\snm{}\ead[label=e2]{david.rios@icmat.es}}
\and
\author[C]{\fnms{P. ANGULO}~\snm{}\ead[label=e3]{pablo.angulo@upm.es}}

\address[A]{Universidad Autónoma de Madrid (UAM), Escuela de Doctorado, Madrid, Spain
\printead[presep={,\ }]{e1}}

\address[B]{Inst. Math. Sciences, Spanish Nat. Research Council, Madrid, Spain
\printead[presep={,\ }]{e2}}

\address[C]{Universidad Politécnica de Madrid (UPM), ETSI Navales, Madrid, Spain\printead[presep={,\ }]{e3}}
\end{aug}

\begin{abstract}
The Plackett–Luce model is widely used to deal with probabilities in discrete choice settings. This paper introduces a novel two-level Plackett–Luce model combined with a multinomial logistic scheme that provides the basis for the route choice module in a smart mobility platform. For this, we develop Bayesian inference and prediction mechanisms to capture consumers’ preferences for personalized route recommendations. The model is empirically tested, allowing for refinements and discussion of its applicability. We also illustrate its practical relevance through several use cases, including relevant route selection, coordinated car pooling, incentive design and synthetic data generation.
\end{abstract}


\end{frontmatter}

\section{Introduction and Background}
\label{sec:introduction}

This paper presents a preference-based model for personalized route choice prediction integrated into an AI-driven, user-centric smart-mobility platform. The platform aims to improve the efficiency and sustainability of transport  systems by accounting for traffic flows, emission reduction, and enhanced accessibility. Among the existing functionalities, it supports integrating multiple transport  modes by coordinating car-pooling with public transport  as a complementary mobility solution. The platform also aims to improve mobility in rural areas, where it may play a key role in ensuring that citizens have access to more reliable transport  options. Overall, the platform promotes route allocation to optimize transport , while explicitly accounting for vulnerable groups to foster inclusiveness. By encouraging eco-friendly travel behavior and leveraging intelligent data analysis for real-time decision support, the platform contributes to smarter and greener urban environments in line with the 11th Sustainable Development Goal.\footnote{\url{https://sdgs.un.org/goals}}

AI and transport  have evolved together for decades. Early work leveraged techniques such as genetic algorithms~\citep{holland1992adaptation} for transport  problems, including bus route optimization and mobility flow prediction~\citep{bielli2002genetic}. Additionally, with the emergence of neural networks, more complex applications appeared, for example in short-term traffic flow prediction~\citep{dougherty1997short}. Subsequently, other widely used statistical techniques, including XGBoost, Support Vector Machines, Naive Bayes, and Random Forests, have been employed in areas such as transport  mode detection~\citep{hasan2022transportation, enev2016automobile}, as well as in modeling the use of more modern transport  options, such as electric scooters, using neural network and random forest models~\citep{zhao2022machine}. This intersection has continued to develop rapidly, and certain advances enable the application of cutting-edge approaches to mobility challenges. For instance, network routing in dynamic environments has also been studied from a statistical decision-making perspective in~\cite{singpurwalla2011routing}. More recently, the use of Large Language Models in intelligent transport  systems has been surveyed in~\cite{shoaib2023survey}, while physics-informed neural networks and generative AI methods in transport  contexts are discussed in~\cite{da2025generative}.

Against this broader background, the model proposed in this paper targets a central component of user-centric mobility services: representing users' preferences from observed route choices. Related contributions in the literature on preference and choice modeling include unfolding-type models, such as~\cite{lei2025unfolding}, which capture latent preference structures from choice data in political settings, and tourism recommender systems, as in~\cite{almomani2023choiceTRS}, which emphasize interpretable preference learning for personalized recommendations. Our approach relies on the Plackett--Luce (PL) framework to model route-choice behavior at the individual level. PL is a widely used generalization of the \cite{bradley1952rank} model, providing a principled mechanism for representing and predicting discrete choices among competing routes. Several other generalized forms of these ranking and choice models have been studied extensively. For instance,~\cite{plackett1975analysis} presents a formulation based on permutations to explain preference orderings, while~\cite{hunter2004mm} explores estimation through a minimization--maximization framework. A related theoretical foundation is provided by~\cite{luce1959individual}, who introduces a choice model grounded in the axiom of independence of irrelevant alternatives, underpinning much of the probabilistic structure used in ranking models. The reasons behind the ubiquity of generalized Bradley--Terry models is explained in ~\cite{hamilton2025manyways2BT}, while a comprehensive treatment of inference for those models is provided by~\cite{caron2012efficient}, where the authors discuss several extensions, including the Plackett--Luce (PL) model, and offer valuable insights into Bayesian methods and practical applications. For a broad and in-depth overview of discrete choice models see~\cite{train2009discrete}.

The overall idea that we pursue is summarized as follows. By representing the route that each user is most likely to choose, individual behavior can be captured and subsequently integrated into a broader decision-making pipeline underlying the incumbent smart mobility platform. In practice, the resulting preference model is used to rank available routes, assign priorities among alternatives, and allocate users accordingly, particularly in settings where demand and capacity must be balanced. Illustrative use cases include coordinating car-pooling with public transport  as a complementary mobility solution, as well as improving mobility in rural areas where ensuring access to reliable transport  options is especially critical. 

Delivering personalized mobility services requires a modeling framework that accounts for individual characteristics to generate tailored alternatives. Accordingly, our main objective is to develop a model capable of representing and describing users' route-choice behavior while accounting for personal attributes. Building on this idea, computational tools are developed to process preference data derived from the routes presented to users, learn sequentially from observed and personalized choices, and predict future decisions conditional on user characteristics. Moreover, the proposed approach is designed to integrate seamlessly in a mobility platform where individualization improves both service quality and system performance.

Methodologically, the proposed PL model is embedded within a two-level multinomial logit (MNL) structure. Bayesian perspectives on MNL inference and route-choice applications are discussed in~\cite{fisher2022bayesian, zellner1983bayesian, fruhwirth2012bayesian} and~\cite{washington2009bayesian}. Going further by incorporating individual attributes through the PL model, our framework attains a higher level of personalization and, consequently, improves flexibility. Moreover, existing work on PL models contains inference approaches based on Expectation-Maximization or maximum-likelihood procedures~\citep{hunter2004mm, francis2010heterogeneity, boonstra2020consensus}, as well as Bayesian formulations that exploit closed-form posterior expressions~\citep{caron2012efficient}. In our case, the added complexity of the proposed model rules out such closed-form updates, and therefore we relay on inference via MCMC, using the No-U-Turn Sampler (NUTS) \citep{hoffman2014no}. To summarize our contributions:
\begin{itemize}
\item We develop a model built on two key components: a PL model and a MNL regression model. This combination allows us to represent individual attributes and contextual factors to describe and predict route choices. The resulting framework is both interpretable and flexible: unlike modern black-box approaches such as deep neural networks and other highly parameterized machine learning models, it retains a clear parametric interpretation, while also being adaptable to different datasets depending on the available information.
\item We propose a Bayesian formulation that supports sequential updating as new data arrive, enabling the model to adapt over time by refining the underlying probability distributions. Additionally, with certain choices such as prior approximations or filtering steps, the resulting procedure
 remains both scalable and efficient.
\item Finally, motivated by its deployment in a smart mobility platform, we demonstrate the practical relevance of the model through four use cases: personalized route selection, coordinated car pooling allocation, personalized incentive design, and synthetic data generation.
\end{itemize}

The remainder of the paper is organized as follows. After this introduction, Sections~2 presents the available data. Then, Sections~3 and~4 
 respectively present the proposed models and inference methods, together with a performance analysis. Section~5 presents use cases and comparisons and Section~6 concludes and outlines directions for future improvement.

\section{Data}
\label{section:data}

The preference models developed within the mobility platform rely on data coming from two main user interactions.
The first one takes place upon registration.
Besides the usual personal information, the user is required to declare:
    their {\em age}, specifically arranged in three groups: young ($<25$), active ($25$--$65$), and retired ($>65$), corresponding to age groups with different economic behavior, different discounts on public transport or lower tariffs;
        their {\em postcode}, used to enrich the available data with socio-economic information, following the 
        procedure described in \cite{corrales2024colorectal}; their   
       {\em disability level}, with three groups: none, moderate, and severe, used to better facilitate more adapted solutions for people who require them. The second one takes place upon trip request. During such interaction, among other information, the
       user declares their earliest starting time and latest arrival time to a desired destination. This 
       is used to to calculate an estimate of the time available to reach the destination, used as a proxy for trip urgency;
        and to make a link to a weather forecast service to construct a dichotomous variable indicating whether rain or adverse weather conditions are forecast at travel time.

 Besides, upon requesting a trip, the platform queries OpenTripPlanner (OTP2) to compute \(K\) potential routes satisfying the requested origin, destination and temporal constraints. OTP2 relies on a multi-criteria routing framework that returns Pareto-optimal itineraries with respect to time and cost.\footnote{OTP2 can be configured with different optimization criteria and penalties.  Here, we refer specifically to the configuration deployed in our server. For further detail, consult \url{https://docs.opentripplanner.org/}}\,\footnote{An itinerary is said to be Pareto-optimal if no other feasible itinerary exists that is at least as good in all considered criteria and strictly better in at least one of them.} Once the routes are computed, we also extract their associated walking times, so that each trip is described by three attributes: travel time, cost, and requested walking time. Finally, when completing the interaction, the user chooses one among the proposed $K$ routes and such preference is registered.

As a consequence, a typical observation in our dataset will be of the form
\[
(z_1,\ldots,z_N,\mathbf{r}_1,\ldots,\mathbf{r}_K,\mathrm{choice}),
\qquad
\mathbf{r}_i=(t_i,c_i,{t_w}_i), \quad i=1,\ldots,K,
\]
where \(z_1,\ldots,z_N\) denote the user features, including both registration information and trip-specific contextual variables, and \(\mathbf{r}_1,\ldots,\mathbf{r}_K\) are the displayed potential routes. Based on discussions with transportation and HCI experts, we set \(K=8\), so that each observation contains information on eight potential routes.

\begin{figure}[t]
    \centering
    \includegraphics[width=0.85\linewidth]{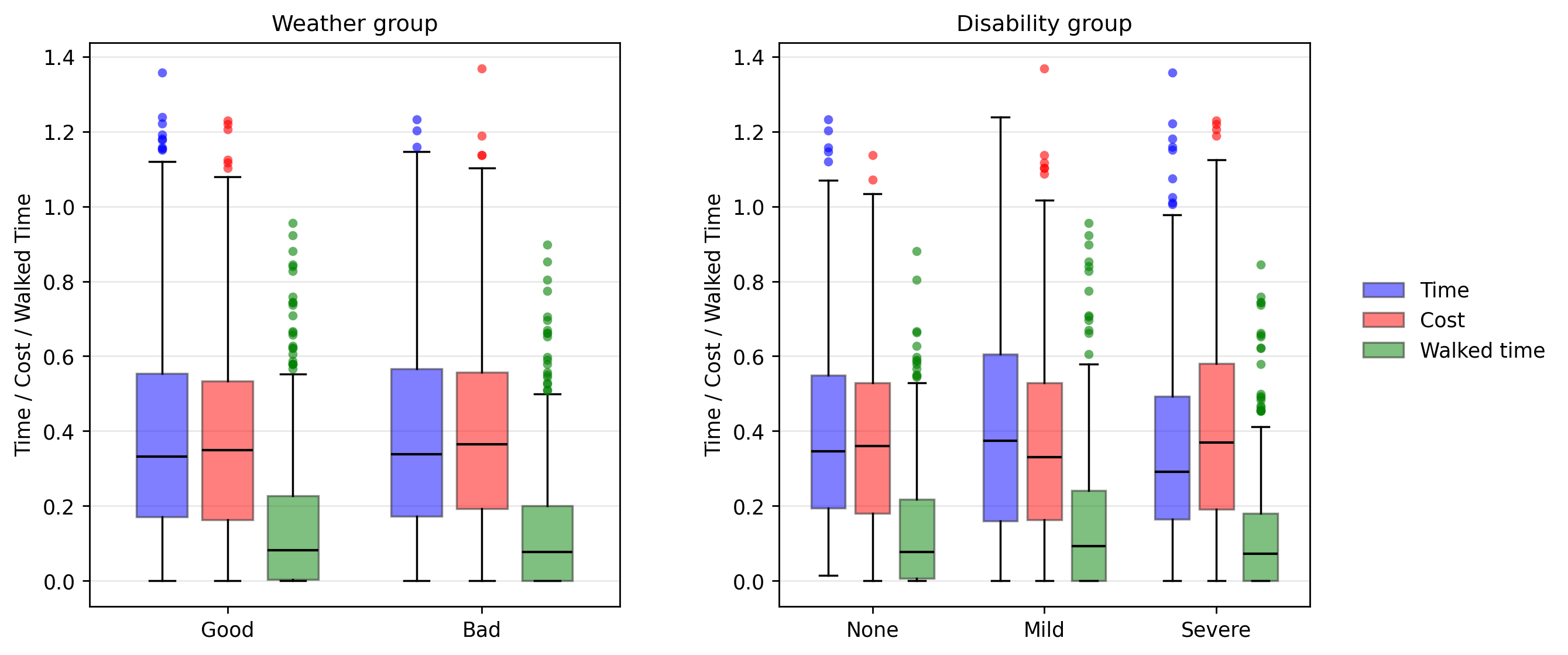}
\caption{Boxplots of scaled features of chosen routes---travel time (\textit{blue}), cost (\textit{red}), and walking time (\textit{green})---across user grouped by weather conditions and disability level}
\label{fig:EDA}
\end{figure}

The exploratory data analysis in \autoref{fig:EDA} suggests the presence of several behavioral patterns. For instance, users with belonging to severe disability groups tend to exhibit slightly lower walking times, as reflected in the upper-tail observations. Similarly, under adverse weather conditions users appear to choose routes with shorter travel and walking times. These effects are not especially evident and are mainly perceived in the distribution tails, reinforcing the idea that route choice is driven by the joint influence of multiple contextual factors rather than by any single variable alone. This motivates the model developed here,  designed to account for the factors simultaneously.

The volume of data generated depends on factors such as the location, the platform’s local adoption, or seasonal patterns.
 In any case, data is captured continuously, but processed nightly to refresh the models.

\section{A Plackett--Luce based preference model}
\label{section: Prefernce Model}

Our prime objective is to develop a model capable of predicting which route a user is most likely to choose, based on personal features and contextual factors, such as whether the user is in a hurry or their economic status, taking into account the route's specific attributes, mainly travel time, monetary cost, and walking time. The goal is therefore to construct a model that takes these parameters as inputs and outputs the probability of choosing each option in a group of routes offered to a user.

The model is formulated using a framework that is capable of capturing individual choices, enlarging the PL model~\citep{plackett1975analysis,luce1959individual}. Following its notation, the probability of route $ l_0 $ being chosen from a set of $ K $   of them $ \{1,2, \dots, K\} $ will be assumed to be given by
\begin{equation*}
P(l_0 \succ \{1, \dots, l_{0}-1, l_{0}+1, \dots, K\}) = \frac{\lambda_{l_0}}{\sum_{l=1}^{K} \lambda_l},
\label{eq:pl_single_choice}
\end{equation*}
where $\lambda_l>0$ represents the \textit{worth} or \textit{value} of the $l$-th route. This expression corresponds to the first term of the product in the PL model in \cite{caron2012efficient}, reflecting a scenario in which only the top-ranked item is relevant and available. The $\lambda_l$ \textit{values} associated with the routes need to be assessed. Assume that the $M$ attributes of the $l$-th route are collected in a vector $\mathbf{r}_l := (r_l^1,\ldots,r_l^M)$, where $r_l^j$ are different attributes, such as travel time or monetary cost. In our specific case, 
  under relatively weak conditions, they are aggregated with a multi-attribute additive value function $v$ \citep{french2000statistical}, through
\begin{equation*}
v(\mathbf{r}_l) = - \sum_{i=1}^M w_i\cdot r_l^i,
\label{eq:additive_value_function}
\end{equation*}
with $w_i \geq 0$ for all $i=1,\ldots,M$ and $\sum_{i=1}^M w_i= 1$. The negative sign is introduced for better interpretability: a higher value of $w_i$ reflects a stronger preference for lower $r^i$ values, where all performance attributes are assumed to be decreasing (less is more preferred).\footnote{In order to compute $v(r_1,\ldots,r_M)$, variables are first standardized so that the additive expression is unit-free.} Moreover, the adopted final expression based on this additive value function will be
\begin{equation*}
    \lambda_l = e^{v(\mathbf{r}_l)} = e^{-\left( \sum_{i=1}^M w_i\cdot r_l^i \right)},
    \label{eq:lambdadefinition}
\end{equation*}
which entails that $\lambda_l\geq 0$. Then, given a set of $K$ routes, characterized by their vector $\mathbf{r}_l$ for $l=1,\ldots,K$, the assumed probability of the $l_0$-th element being chosen will be
\begin{equation}
\Pr \left(l_0 \succ \{1, \dots, l_{0}-1, l_{0}+1, \dots, i_k\}\right) =
\frac{\exp\left( v(\mathbf{r}_{l_0}) \right)}{\sum_{l=1}^{K} \exp\left( v(\mathbf{r}_{l}) \right)},
\label{eq:pl_softmax}
\end{equation}
which can be regarded as a softmax over the values $v(\mathbf{r}_{i})$ for $i=1,\ldots,K$.
To this end, the weights $w_j$ are readily \textit{interpretable} by considering the \textit{odds} of choosing option $l_k$ over $l_j$,
\begin{equation}
\text{odds} \left(l_k \succ l_j\right) =
\frac{\exp\left( v(\mathbf{r}_{l_k}) \right)}{\exp\left( v(\mathbf{r}_{l_j}) \right)}=
\prod_i\exp(-w_i(r_{l_k}^i-r_{l_j}^i)).
\label{eq:pl_softmax2}
\end{equation}
Thus, $w_i$ is interpreted as the logarithm of the odds that route $l_j$ is chosen over $l_k$ if they differ only in one unit of the $i$-th criteria.

To obtain personalized responses, the next step is to model the weights $\mathbf{w}$ as depending on both contextual and individual features associated with each user interaction. Suppose, these features are collected in a vector $\mathbf{z}:=(z_1,\ldots,z_N)$. Additionally, for each weight $w_j$, $j=1,\dots,M$, a parameter vector $\mathbf{a}^j$ is introduced and the weights are linked to $\mathbf{z}$ through a MNL model
\begin{equation}
\mathbf{(a}^j)^\top \mathbf{z} = \sum_{k=1}^N a_k^j z_k, \quad j = 1, \ldots, M, \qquad
w_i = \frac{\exp(\mathbf{(a}^i)^\top \mathbf{z})}{\sum_{j=1}^M \exp(\mathbf{(a}^j)^\top \mathbf{z})} = w_i(\mathbf{a, z}),
\label{eq:softmax_weights}
\end{equation}
where $\mathbf{a^1}$ is set equal to $\mathbf{0}$ to mitigate identifiability issues. Specifically, if $\mathbf{a} = (\mathbf{a}^1, \ldots, \mathbf{a}^M)$ is a solution, then $\mathbf{a}^* = (\mathbf{a}^1 - \boldsymbol{\alpha},  \ldots, \mathbf{a}^M- \boldsymbol{\alpha})$ would also be consistent.\footnote{Another possible approach to solve identifiability fixes a different weight, or use a more general constraint such as forcing  the sum of parameters to be constant, $\sum_j \mathbf{a}^j = c$. However, in order to maintain simplicity, applying $\mathbf{a}^1 = 0$ reduces the number of parameters involved, thereby decreasing the dimensionality of the space over which the gradient is computed and, consequently, reducing the computational cost.} Moreover, the notation $\mathbf{w}(\mathbf{a,z}) = (w_1(\mathbf{a,z}), \ldots, w_M(\mathbf{a,z}))$ is used henceforth.\footnote{From this point onward, the notation $\mathbf{a}$ will be used equivalently to refer to the unknown parameters $(\mathbf{a}^2, \ldots,\mathbf{a}^M)$, or equivalently $(\mathbf{a}^1 = 0, \ldots, \mathbf{a}^M)$, even though the first component is fixed and no inference about it is to be done.} 

The final modeling step consists of selecting a route. Given the probabilities computed for each of the $K$ routes presented to a user, the chosen route is defined as the one with the highest predicted probability of being chosen. However, the computed probabilities still represent our estimate of the uncertain choice that the user will make.

\begin{figure}[t]
    \centering
    \begin{tikzpicture}[node distance=1.cm]
    
        \node (A) [block, align=center] {Model Parameters \\ $\mathbf{a}=(\mathbf{a}^1=\mathbf{0},\ldots,\mathbf{a}^M)$ };
        \node (B) [block, right=0.2cm of A, align=center] {User and Context Information \\ $\mathbf{z}=(z_1,\ldots,z_N)$};
        \node (C) [block, right=0.2cm of B, align=center] {Route's Information \\ $\{\mathbf{r}_l=(r_l^1,\ldots,r_l^M)\}_{l=1}^K$};
        
        \begin{pgfonlayer}{background}
            \node[
                draw=black,
                thick,
                fill=gray!10,
                rounded corners,
                fit=(B)(C),
                inner xsep=0.1cm,   
                inner ysep=0.2cm   
            ] {};
        \end{pgfonlayer}

        \node (D) [block, below=1.0cm of B, align = center, text width = 5.8cm] {Weights \\ Multinomial Logit Regression \\ $w_i = \frac{\exp({\mathbf{(a}^i)^\top \mathbf{z}})}{\sum_j \exp({\mathbf{(a}^j)^\top \mathbf{z}})}$ for $i=1,\ldots,M$.};

        \node (E) [block, below=1.0cm of D, align=center, text width=9.5cm] {%
            Plackett--Luce Model\\[2pt]
            {\small
            $\displaystyle
            \left\{
            \Pr\!\left(l \succ \{1,\dots,l-1,l+1,\dots,K\}\right)
            =
            \frac{\exp\!\big(v(\mathbf{r}_{l})\big)}
            {\sum_{j=1}^{K}\exp\!\big(v(\mathbf{r}_{j})\big)}
            \right\}_{l=1}^{K}
            $}
        };

        \node (F) [block, below=0.8cm of E, align = center] {User's Choice: $\textcolor{blue!80!black}{\mathbf{l_0}}$ };;
        
        \begin{pgfonlayer}{background}
            \node[
                draw=black,
                thick,
                fill=gray!10,
                rounded corners,
                fit=(F),
                inner xsep=0.2cm,   
                inner ysep=0.2cm
                ]{};
        \end{pgfonlayer}

        \draw [arrow] (A.south) --++ (0,-0.52) -| (D.north);
        \draw [arrow] (B.south) -- (D.north);
        \draw [arrow] (C.south) --++ (0,-3.35) -| (E.north);
        \draw [arrow] (D.south) -- (E.north);
        \draw [arrow] (E.south) -- (F.north);
    \end{tikzpicture}
    \caption{Diagram of the two-level Plackett-Luce preference model. Boxes in gray designate observed variables.}
    \label{fig:Modeldiagram}
\end{figure}
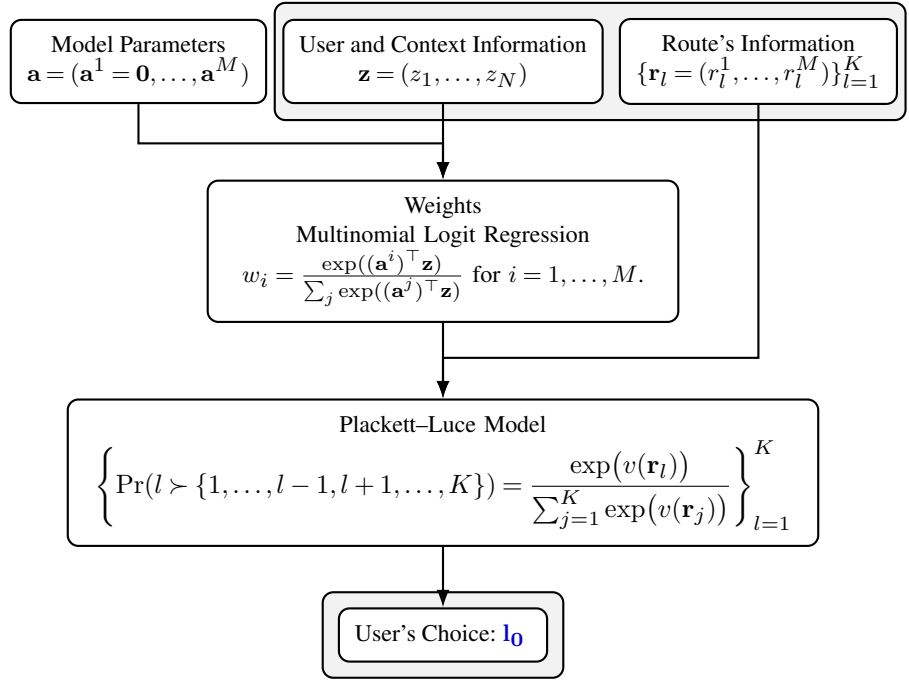

\autoref{fig:Modeldiagram} shows the scheme of the proposed preference model, which incidentally can be interpreted as a neural network architecture with $M\cdot K + N$ input nodes; $M$ hidden nodes in the first layer applying softmax transformations to compute the weights $\mathbf{w}$ from $\mathbf{z}$; $K$ linear hidden nodes in a second layer; and, a final softmax output node.

\textbf{Specification of the preference model.} The inference procedure and use cases presented in the next sections require an explicit definition of the route characterization and the user-feature set based on the available data as discussed in Section \ref{section:data}.

\textit{Route attributes.}
Each route is characterized by its total duration \(t\), its cost \(c\), and its entailed walking time \(t_w\). Hence \(M=3\) and, for route \(l\),
\(\mathbf{r}_l=\bigl(r_l^1,r_l^2,r_l^3\bigr)=\bigl(t_l,\;c_l,\;{t_w}_{l}\bigr)\).
Accordingly, there are three preference weights
\(\mathbf{w}=(w_1,w_2,w_3)\). When there are \(K=8\) possible routes, the compact notation  \(p(\mathbf{w},\mathbf{c},\mathbf{t},\mathbf{t}_w)\) of the choice will be used used,
with \(\mathbf{c}=(c_1,\ldots,c_k)\), \(\mathbf{t}=(t_1,\ldots,t_k)\), and
\(\mathbf{t}_w=({t_w}_{1},\ldots,{t_w}_{k})\).

\textit{User and context features.} \(\mathbf{z}=(z_1,\ldots,z_N)\) are presented in the following order:
\begin{itemize}
    \item \(z_1\): user age according to three groups (young \(<25\), active \(25\)--\(65\), retired \(>65\)), one-hot encoded as \(z_1=(z_{11},z_{12},z_{13})\) with \(z_{1k}\in\{0,1\}\) and \(\sum_{k=1}^3 z_{1k}=1\).
    \item \(z_2\): user socio-economic level derived from postcode (increasing with socio-economic status). 
    \item \(z_3\): weather indicator ($1$ if rain/bad conditions are forecast at travel time, $0$ otherwise), assuming reliable short-term forecasts available.
    \item \(z_4\): estimation of time available to reach destination, used as a proxy for urgency.
    \item \(z_5\): user disability level: 0 none, 1 moderate, 2 severe.
\end{itemize}
Consequently, for each weight \(w_j \) with \( j\in\{1,2,3\}\), the assigned feature vector $\mathbf{a}^j$ will be defined as \begin{equation}
    \mathbf{a}^j = \left(a_{11}^j, \,a_{12}^j, \,a_{13}^j, \,a_{2}^j, \,a_{3}^j, \,a_{4}^j, \,a_{5}^j\right).
    \label{eq:Parameterstructure}
\end{equation}
with $\mathbf{a}^1=\mathbf{0}$.

\section{Inference and prediction}
\label{sec: Proposed Inference Algorithms}

This section presents inference and prediction procedures for the proposed model (\autoref{eq:pl_softmax}, \autoref{eq:softmax_weights}) reflected in Figure 2. Our main objective is to infer the posterior distribution of the weight parameters \(\mathbf{a}\), given the observed data and prior information, and use it for prediction and decision support, as illustrated by the use cases in Section~5. On day \(l\), the system stores the data \(d_l\), containing the interaction and preference data recorded on that day. The cumulative dataset evolves as \(D_l = D_{l-1} \cup d_l\), where \(D_l\) denotes all observations collected up to day \(l\), for \(l \geq 1\). Normal priors with mean zero and unit variance are assigned to the model parameters, reflecting the lack of specific prior information and the fact that, as data accumulate, the likelihood quickly dominates the prior.

Two modeling scenarios are considered. In the first one, referred to as the \textit{static case}, inference is performed using all data available up to day \(l\), namely \(D_l\). This approach is mainly useful for interpretation and comparison, but it is not suitable for production, since it requires rerunning inference each day and storing the full dataset. To overcome these limitations, we also developed several \textit{dynamic} approaches and report the best-performing alternative. In the dynamic setting, the model is updated sequentially overnight as new data arrive, providing an efficient and scalable inference procedure over time. This allows updated parameter estimates to be available for use the following day.

Posterior inference is carried out using NUTS, which is well suited to the moderate dimensionality of the model. The sampler is implemented in Python with the \texttt{NumPyro} library, built on top of \texttt{JAX} for efficient tensor operations and automatic differentiation.

\subsection{Static model}
\label{subsec: Static Model}

The procedure considered focuses on the case following the $l$-th day of data collection. Assume a prior distribution $\pi_0(\mathbf{a})$ is available, representing initial beliefs about the parameters $\mathbf{a}$ before observing the data. The prior is updated using Bayes' formula,
\begin{equation*}
\pi(\mathbf{a} \mid D_l) \propto 
\left[ \prod_{i=1}^{N_{\text{obs}}} 
p\left( \mathbf{w}(\mathbf{a} z^i), \mathbf{c}^i, \mathbf{t}^i, \mathbf{tw}^i \right) \right] 
\pi_0(\mathbf{a})
\label{eq:posterior_update}
\end{equation*}
where $p\big( \mathbf{w}(\mathbf{a z^i}), \mathbf{c}^i, \mathbf{t}^i, \mathbf{t_w}^i \big) :=  \Pr_{\,(i-th\,\, observation)} \left( (t_j, c_j, t_{w_j}) \succeq \{(t_l, c_l, t_{w_l})\}_{l\neq j} \right)$ represents the probability associated with the choice of route $j$ by the user in the $i$-th observation and $N_{\text{obs}}$ is the number of user interactions in $D_l$. A scheme for its implementation is in \autoref{alg:NUTS Static Version}, where the dimensions are adapted to our specification in~\autoref{section: Prefernce Model}

\begin{algorithm}[ht]
\caption{\textbf{Static Model}}
\label{alg:NUTS Static Version}

\begin{algorithmic}[1]
\Statex \hspace{-0.5cm}\textbf{Input:} Dataset of $N_{\text{obs}}$ observations; MCMC hyperparameters $n_{\text{burnout}}$, $n_{\text{samples}}$.
\Statex \hspace{-0.5cm}\textbf{Output:} Posterior samples $\{\mathbf{a}\}_{i=1}^{n_{\text{sample}}}$.
\vspace{-0.2cm}
\Statex \hspace{-0.6cm} \rule{1.05\linewidth}{0.4pt}

\State Transform data to the input shape, $\mathbf{Z} \in \mathbb{R}^{N_{\text{obs}} \times 7}$, $\mathbf{T}, \mathbf{C}, \mathbf{TW} \in \mathbb{R}^{N_{\text{obs}} \times K}$ and $\mathbf{W} \in \{1, \dots, K=8\}^{N_{\text{obs}}}$.

\State Define \texttt{model()}:
\Statex \quad 2.1: Define priors: $a^2, a^3 \sim \mathcal{N}(0, I_7)$. ($a^1=0$)
\Statex \quad 2.2: \textbf{for} each observation \textbf{do}
\Statex \quad \quad \quad \quad  Compute weights: $w_j = \frac{\exp(a^j z)}{\sum_{k=1}^3 \exp(a^k z)} \quad \text{for } j = 1, 2, 3$
\Statex \quad \quad \quad \quad Compute values: $v_j = -w_1 t_j - w_2 c_j - w_3 tw_j$
\Statex \quad \quad \quad \quad Compute probabilities: $p_j = \frac{\exp(v_j)}{\sum_{l=1}^K \exp(v_l)}$
\Statex \quad \quad \quad  \quad Define likelihood: $W_i \sim \texttt{Categorical}(\mathbf{p})$
\Statex \quad \quad end \textbf{for}
\State \textbf{Run inference:} initialize \texttt{NUTS()} kernel, run \texttt{mcmc.run()}, extract posterior with \texttt{mcmc.get\_samples()}
\end{algorithmic}
\end{algorithm}

\subsection{Dynamic model}
\label{subsec: Simple Dynamic Model}

\vspace{-0.2cm}
As mentioned, the dynamic nature of the incoming data entails a sequential processing approach where the posterior distribution obtained on day $(l-1)$, serves as prior for day $l$. This not only avoids storing large amounts of data with very old observations, but also enhances computational efficiency. In this way, the posterior is incrementally updated using Bayes' formula as new data becomes available. As standard, given a dataset $d_l$ with $N_l$ new user interactions on day $l$, the posterior distribution is updated to
\begin{equation*}
\log \pi(\mathbf{a} \mid D_l) \propto
\sum_{i=1}^{N_l} 
\log p\left( \mathbf{w}(\mathbf{a} z^i), \mathbf{c}^i, \mathbf{t}^i, \mathbf{tw}^i \right) 
+ \log \pi(\mathbf{a} \mid D_{l-1}) ,
\label{eq:dynamic_update}
\end{equation*}
where the first term on the right-hand side corresponds to the likelihood of the newly observed data, and the second term is the log-posterior from the previous step. With this basic scheme in mind, several considerations are made in order to build the final dynamic version.

\textbf{I. Prior distribution.} A straightforward approach, as described in \cite{insua2012bayesian}, is to use the posterior samples as a prior discrete distribution. However, HMC-methods rely on gradient-based sampling. Another motivation for this continuous approximation is rooted in the structure of Bayesian inference itself. The posterior is proportional to the prior, implying that any region where the prior is zero will also result in a zero posterior probability. Consequently, using discrete samples alone as a prior distribution, would prevent the posterior from evolving beyond those specific sample points, a limitation that may bias inference. To address this, the prior is approximated with a continuous mixture of Gaussian densities centered at the posterior samples making the prior compatible with the HMC framework. Specifically, we employ, 
\begin{equation*}
    \pi(\mathbf{a} \mid D_{l-1}) = \frac{1}{n_{\text{sample}}} \sum_{i=1}^{n_{\text{sample}}} \frac{1}{(2\pi)^{\frac{d}{2}} |\mathbf{\Sigma}|^{\frac{1}{2}}} \exp\left(-\frac{1}{2} (\mathbf{a} - \mathbf{a}_i)^\top \mathbf{\Sigma}^{-1} (\mathbf{a} - \mathbf{a}_i) \right),
    \label{sumofGaussiansformula}
\end{equation*}
where \(d\) denotes the dimensionality of \(\mathbf{a}\), and \(n_{\text{sample}}\) is the number of posterior samples available. 

The covariance matrix \(\Sigma\) is critical in this formulation. Large $\Sigma_{ii}$ values within the diagonal allow the model to explore regions of the parameter space farther from the original samples, while a smaller value keeps the density concentrated near the previous sample points, increasing stability. However, setting the variance too low may cause the sampling process to be overly sensitive to the initial point in the Hamiltonian trajectory, potentially hindering posterior exploration. Furthermore, the choice of \(\Sigma\) depends on $n_{\text{sample}}$ and $d$: a larger number of samples generally supports a lower variance, while higher dimensional spaces typically require even smaller variances, as the distances between points tend to increase with the dimension.

While this formulation is general, the full covariance matrix \(\mathbf{\Sigma}\) introduces a significant computational overhead, particularly because gradients must be evaluated at each step of the MCMC algorithm. In order to simplify computation, the covariance matrix is taken to be diagonal, $\Sigma \approx \text{diag}(\mathbf{\Sigma})$. Moreover, and to define $\text{diag}(\mathbf{\Sigma})$ in a suitable manner for gradient-based sampling, a plug-in estimator is used. Based on \cite{chacon2018multivariate}, this diagonal bandwidth matrix for the first order derivative density estimation is
\begin{equation*}
    \text{diag}(\mathbf{\Sigma}) \approx \left(\frac{4\cdot n_{\text{sample}}}{d + 4 }\right)^{\frac{2}{d + 6} } \text{diag}(S),
\end{equation*}
where $S$ denotes the sample covariance matrix.

\textbf{II. Particle filtering.} Another improvement to the proposed sequential prior-posterior scheme is inspired by filtering strategies commonly employed in SMC, see e.g. \cite{doucet2000sequential, burda2023hamiltonian}. The idea is to refine the posterior sample from iteration \( (l - 1) \) using the newly available data at iteration  \( l \)  before constructing the prior. The procedure is composed of these steps:
\begin{enumerate}
    \item[i.] \textit{Weight Calculation.} For each particle \(\{\mathbf{a}_j\}_{j=1}^{n_{\text{sample}}}\), a log-weight is computed from the data loglikelihood \(
        \xi_j = \sum_{i=1}^{N_l} \log p(\mathbf{w}(\mathbf{a}_j \mathbf{z}^i), \mathbf{c}^i, \mathbf{t}^i, \mathbf{tw}^i)
    \) \footnote{Using log-likelihoods avoids numerical underflow.}.
    \item[ii.] \textit{Weight Normalization.} Define shifted log-weights \(\xi_j^* = \xi_j - m\) where \(m = \text{median}((\xi_k)_{k=1}^{n_{\text{sample}}})\). Normalized weights are then obtained through
    \(\omega_j = {e^{\xi_j^*}}/{\sum_{k=1}^{n_{\text{sample}}} e^{\xi_k^*}}\), corresponding to a numerically stable softmax transformation.

    \item[iii.] \textit{Deterministic Resampling.} Influential particles are copied deterministically: each particle \(j\) is included \(\lfloor \omega_j\cdot n_{\text{sample}} \rfloor\) times in the new sample.

    \item[iv.] \textit{Probabilistic Resampling.} The remaining \(n^* = n_{\text{sample}} - \sum_{j=1}^{n_{\text{sample}}} \lfloor \omega_j \cdot n_{\text{sample}} \rfloor\) particles are drawn with probabilities proportional to the residuals
    \(\delta_j = \omega_j\cdot n_{\text{sample}} - \lfloor \omega_j \cdot n_{\text{sample}}, \rfloor\) after normalizing \((\delta_j)_{j=1}^{n_{\text{sample}}}\).

\end{enumerate}
As a consequence, a refined particle sample is obtained that incorporates the most recent data while preserving the structure of the previous posterior, and it serves as prior for the next iteration.

\textbf{III. Weighted likelihood.} To handle limited data on a certain day $l$ and accommodate possible temporal shifts in behavior, we define a weighted likelihood with observation-specific weights. These weights regulate each observation's contribution, assigning greater influence to data points that are more contextually relevant to the setting where the inferred parameters will be used. As an example, travel patterns often differ between weekdays and weekends, or time constraints are typically more binding on weekdays. The resulting likelihood therefore allows heterogeneous influence across observations rather than assuming all data are equally informative.

Let \(N_{\text{obs}}=\sum_{l=0}^L N_l\) denote the total number of observations recorded up to a given day $L$, regardless of when they were collected or whether they have already been used in previous inference steps. Each observation \(i\) is assigned a weight \(\eta_i\), \(i = 1, \dots, N_{\text{obs}}\), defined according to a desired criterion. The corresponding weighted likelihood is
\begin{equation*}
    L(\mathbf{a} \mid D_L) = 
    \prod_{i=1}^{N_{\text{obs}}}
    \left(
        p\bigl(\mathbf{w}(\mathbf{a}, \mathbf{z}^i), \mathbf{c}^i, \mathbf{t}^i, \mathbf{tw}^i\bigr)
    \right)^{\eta_i}.
    \label{eq:weighted_likelihood_v3}
\end{equation*}
We propose a simple age-based weighting scheme in which the weight of observation \(i\) depends only on its antiquity:
$\eta_i=\beta^{\alpha_i}$, where \(\beta\in[0,1]\) is a temporal decay factor and \(\alpha_i := l-(k_i+1)\in\mathbb{Z}\) denotes the observation’s age.\footnote{Alternative weighting schemes can be used to reflect recurring patterns, for example
$\eta_i=\beta^{\alpha_i}\Big(\lambda \delta_i+(1-\lambda)(1-\delta_i)\Big),
$ where $\lambda\in[0,1]$ controls the relative importance of weekday versus weekend data, and $\delta_i=1$ if observation $i$ belongs to the same day-type category (weekday/weekend) as the target setting, and $\delta_i=0$ otherwise.}

Although this weighted likelihood incorporates more information and can improve inference accuracy, it also has an obvious drawback: the number of observations used at each iteration is not fixed and may grow over time. This can lead to increased computational cost, since no data are ever entirely removed, only down-weighted. To avoid overwhelming the model with a large and potentially outdated dataset, it is advisable to impose a maximum age threshold. For example, observations with \(l - (k_i + 1) > A_{\text{max}}\) will be discarded and removed from the system. Moreover, one may select a value \(N_{\text{max}}\) and retain only the \(N_{\text{max}}\) observations with the highest weight, that is, those expected to have the greatest influence on the inference results. In this way, inference execution time is controlled, and at each step the number of observations used is set to \(N_{\text{obs}}=\min\{N_{\text{max}},\sum_{l=L-A_{\text{max}}}^L N_l\}\).\footnote{For clarity, this weighted-likelihood approach preserves the idea that today’s posterior is tomorrow’s prior, while still allowing the likelihood to include previously used observations in a controlled manner. This ensures that each day the likelihood is built from a sufficiently large set of observations, even if some observations contribute more than once to the inference process (with appropriate down-weighting).}

Taking into account the sequential prior–posterior updating scheme and the three considerations discussed, the resulting final algorithm employed is as follows. 

\begin{algorithm}[ht]
\caption{\textbf{Dynamic model}}
\label{alg: dynamicversion}

\begin{algorithmic}[1]
\Statex \hspace{-0.5cm}\textbf{Input:} Dataset with $N_{\text{obs}}$ observations and \textbf{weights} $\eta_i$; MCMC hyperparameters: $n_{\text{burnout}}$, $n_{\text{samples}}$; posterior samples of day $l-1$, $\{\mathbf{a}\}_{i=1}^{n_{\text{sample}}}$.

\Statex \hspace{-0.5cm}\textbf{Output:} Posterior samples of day $l$, $\{\mathbf{a}\}_{i=1}^{n_{\text{sample}}}$.
\vspace{-0.2cm}
\Statex \hspace{-0.75cm } \rule{1.05\linewidth}{0.4pt}  
\State Transform data to $\mathbf{Z} \in \mathbb{R}^{N_{\text{obs}} \times 7}$, $\mathbf{T}, \mathbf{C}, \mathbf{TW} \in \mathbb{R}^{N_{\text{obs}} \times k}$ and $\mathbf{W} \in \{1, \dots, k\}^{N_{\text{obs}}}$.

\State \textbf{Sample Filter:}
\Statex \quad 2.1: \textbf{for} $j = 1, \ldots , n_{\text{samples}}$ \textbf{do}
\Statex \quad \quad\quad \quad Compute $\xi_j^* = \xi_j - \text{median}((\xi_k)_{k=1}^{n_{\text{samples}}})$ where $\xi_j = \sum_{i=1}^{N_L} \log p(\mathbf{w}(\mathbf{a}_j, \mathbf{z}^i), \mathbf{c}^i, \mathbf{t}^i, \mathbf{tw}^i)$.
\Statex \quad\quad \quad \quad Compute $\omega_j = {e^{\xi_j^*}}/{\sum_{k=1}^{n_{\text{samples}}} e^{\xi_k^*}}$
\Statex \quad\quad \quad \quad \textbf{if} $\lfloor \omega_j \cdot n_{\text{samples}} \rfloor \geq 1$ \textbf{then} add $\mathbf{a}_j$ to $\{\mathbf{a}^{\text{new}}\}$ $\lfloor \omega_j \cdot n_{\text{samples}} \rfloor$ times \textbf{end if}
\Statex \quad\quad \quad \quad Compute  $\delta_j = \omega_j \cdot n_{\text{samples}} - \lfloor \omega_j \cdot n_{\text{samples}}\rfloor$
\Statex \quad \quad \quad end \textbf{for}

\Statex \quad 2.2: Compute $n^* = n_{\text{samples}} - \sum_{j=1}^{n_{\text{samples}}} \lfloor \omega_j \cdot n_{\text{samples}} \rfloor$
\Statex \quad 2.3: Sample $n^*$ particles from $\{\mathbf{a}_i\}_{i=1}^{n_{\text{samples}}}$ with probs $\propto \delta_j$.

\vspace{0.2cm}

\State Define \texttt{SumofGaussians()} class to sample from prior (\texttt{Input}: $\{\mathbf{a}_i\}_{i=1}^{n_{\text{samples}}}$ from day $l-1$):
\Statex \quad 2.2: Compute $\text{diag}(S)$ with $S = \frac{1}{n_{\text{samples}}} \sum (\mathbf{a} - \bar{\mathbf{a}})(\mathbf{a} - \bar{\mathbf{a}})^\top$
\Statex \quad 2.3: Compute $\hat{H}_{\text{NS},1} = \left(\frac{4n_{\text{samples}}}{d + 4}\right)^{\frac{2}{d + 6}} \text{diag}(S)$
\Statex \quad 2.4: Define \texttt{log-likelihood} centered at $\{\mathbf{a}_i\}_{i=1}^{n_{\text{samples}}}$ with covariance $\hat{H}_{\text{NS},r}$
\vspace{0.2cm} 

\State Define Bayesian \texttt{model()}:
\Statex \quad 3.1: Prior: $\mathbf{a} \sim \texttt{SumofGaussians()}$
\Statex \quad 3.2: \textbf{for} each observation \textbf{do}
\Statex \quad \quad \quad $w_j = \frac{\exp(a^j z)}{\sum_{k=1}^3 \exp(a^k z)}$
\Statex \quad \quad \quad  $p_j = \frac{\exp(v_j)}{\sum_{l=1}^K \exp(v_l)}$ with $v_j = -w_1 t_j - w_2 c_j - w_3 tw_j$.
\Statex \quad \quad \quad $W_i \sim \texttt{Categorical}(\mathbf{p})$
\Statex \quad \quad \quad $likelihood_i \sim \eta_i W_i$
\Statex \quad \quad \quad end \textbf{for}
\vspace{0.2cm}

\State \textbf{Run inference:} initialize \texttt{NUTS()} kernel, run \texttt{mcmc.run()}, extract posterior with \texttt{mcmc.get\_samples()}
\end{algorithmic}
\end{algorithm}
\FloatBarrier
\subsection{Performance analysis}

To evaluate the inference algorithm, a series of experiments were carried out to assess its performance. This evaluation made it possible to identify potential weaknesses and, based on the results, motivate improvements and the considerations proposed.

\textbf{Static model.} A very large number of observations ($200{,}000$) has been used to establish a benchmark for good performance and facilitate interpretation of results.

\begin{figure}[ht]
    \centering
    \includegraphics[width=0.95\linewidth]{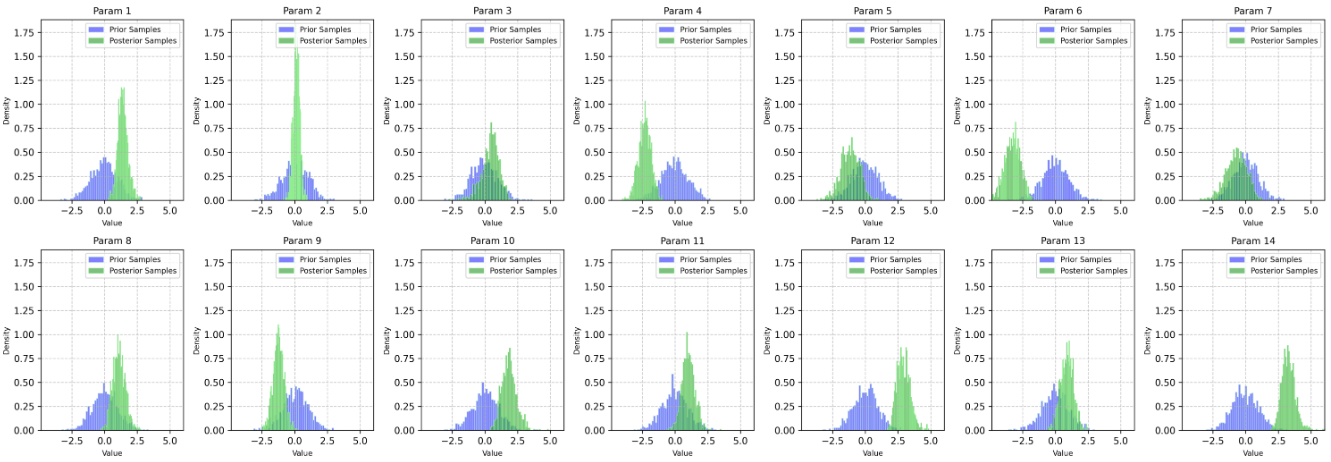}
    \caption[Static algorithm results.]{Posterior distributions (in green) of the 14 model parameters $(\mathbf{a^2},\mathbf{a^3})$, in subplots following the structure in \autoref{eq:Parameterstructure}. Prior distributions in blue for comparison.}
    \label{fig:Results/Static_500KObs_36min.png}
\end{figure}
As \autoref{fig:Results/Static_500KObs_36min.png} shows, the parameters appear to converge to stable values while preserving the normality structure assumed in the prior. As the posterior distributions are Gaussian shaped, the MAP estimate and the posterior mean coincide. Therefore, the vector of posterior means can be safely used as point estimate for \textbf{a}, reported in \autoref{tab:meanpredictedvalues}. The key information to extract from these parameters concerns the hierarchy within the model: parameters with larger magnitudes indicate features that exert a stronger influence on choices. The sign, positive or negative, is less relevant here because we have set the first row of parameters equal to zero. What matters is the relative magnitude of a parameter compared to the other parameters of the same feature variable \(i\), that is, the relative size of \(a_i^1, a_i^2, a_i^3\).

\begin{table}[ht]
\caption{Predicted values of model parameters for static algorithm based on a likelihood built up with 200,000 observations. \textit{(top)} Posterior means; \textit{(bottom)} 90\% credible intervals for posterior means.}
\label{tab:meanpredictedvalues}
\centering
\tiny
\setlength{\tabcolsep}{5pt}
\renewcommand{\arraystretch}{1.10}

\begin{tabular}{l ccccccc}
\toprule
 & $\mathbf{z_{11}}$ & $\mathbf{z_{12}}$ & $\mathbf{z_{13}}$ & $\mathbf{z_2}$ & $\mathbf{z_3}$ & $\mathbf{z_4}$ & $\mathbf{z_5}$ \\
\midrule
$\mathbf{w_2}$ (c)   & 1.402 & 0.135 & 0.446 & -2.315 & -1.155 & -3.263 & -0.716 \\
$\mathbf{w_3}$ (t\_w) & 1.110 & -1.243 & 1.843 & 1.018 & 2.934 & 0.884 & 3.281 \\
\bottomrule
\end{tabular}

\vspace{0.22cm}

\setlength{\tabcolsep}{3.2pt}
\begin{tabular}{l ccccccc}
\toprule
 & $\mathbf{z_{11}}$ & $\mathbf{z_{12}}$ & $\mathbf{z_{13}}$ & $\mathbf{z_2}$ & $\mathbf{z_3}$ & $\mathbf{z_4}$ & $\mathbf{z_5}$ \\
\midrule
$\mathbf{w_2}$ (c)   &
[ 0.760 , 2.033 ] &
[ -0.261 , 0.545 ] &
[ -0.685 , 1.451 ] &
[ -3.127 , -1.579 ] &
[ -2.218 , 0.020 ] &
[ -4.236 , -2.283 ] &
[ -1.983 , 0.575 ] \\
$\mathbf{w_3}$ (t\_w) &
[ 0.297 , 1.816 ] &
[ -1.935 , -0.610 ] &
[ 0.997 , 2.811 ] &
[ 0.280 , 1.866 ] &
[ 2.038 , 3.629 ] &
[ 0.092 , 1.684 ] &
[ 2.531 , 4.155 ] \\
\bottomrule
\end{tabular}
\end{table}

\noindent As an example, consider $z_{12}$, indicating whether an individual belongs to the adult age group. The parameter $a^{3}_{12} = -1.243$, associated with $w_3$ (weight on walked time), is smaller than the corresponding baseline value $0$ for $w_1$ (time). This does not imply that being an adult has a ``negative'' effect on walked time in the usual sense. Instead, since the parameters are exponentiated to obtain the weights, the correct interpretation is that $z_{12}$ reduces the weight placed on walked time more strongly than it reduces the weight placed on time, and differently from cost, whose parameter $a^{2}_{12} = 0.135 > 0$ increases the weight assigned to cost. In other words, for adult individuals, the model shifts relative importance away from walked time towards time and cost, which receive comparatively more weight in the route choice process. Moreover, credible intervals in \autoref{tab:meanpredictedvalues} might be applied for feature selection by excluding covariates $z_j$ whose both associated latent components $(a^2_j, a^3_j)$ are close to zero, since such variables would exert a negligible influence on customer route selection.

Once interpretability has been assessed, a convergence measure is needed to compare performance across scenarios. To this end, we compute the following accuracy metric on a validation set
\begin{equation*}
    \text{Acc} 
    = \frac{1}{N_{\text{val}}}
      \sum_{i=1}^{N_{\text{val}}} 
      \mathbb{I}\!\left(i_0^{(i)} = \arg\max_{j\in\{1,2,\ldots,8\}} \hat{p}_j^{(i)} \right),
\end{equation*}
where \(N_{\text{val}} = 200{,}000\), \(i_0^{(i)}\) is the route actually chosen by the \(i\)-th user, and \(\arg\max_j \hat{p}_j^{(i)}\) is the predicted choice, obtained as the route with highest predicted probability using the posterior mean estimates. Thus, \(\text{Acc}\) measures the proportion of cases in which the model's recommendation matches the observed choice.

For the benchmark simulation with \(N_{\text{obs}}=200{,}000\), the resulting validation accuracy is \(16.56\%\). Although modest in absolute terms, this improves over the uniform baseline of \(1/8=12.5\%\), since each user is presented with eight routes. This result is also consistent with the predicted probabilities, as the top-ranked route typically receives only around \(15\%\) to \(25\%\). On the contrary, this suggests substantial residual variability in user decisions, likely due to contextual or personality features not available in the data presented in \autoref{section:data}.

\begin{figure}[htbp]
    \centering
    \includegraphics[width=0.72\linewidth]{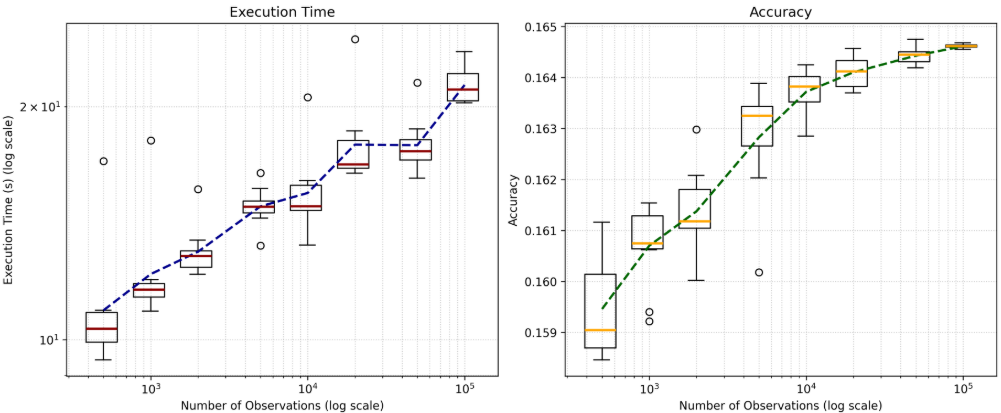}
    \caption[Comparison of static-model performance as a function of the number of observations.]{Overall performance of static model across different likelihood sample sizes and repeated runs. Boxplots summarize variability across runs and dashed lines indicate the mean. (\textit{left}) Log--log plot of computational time versus number of observations used to build likelihood. (\textit{right}) Log--log plot of validation accuracy versus number of observations.}
    \label{fig:Results/Static_comparation.png}
\end{figure}

Using the previously defined performance and convergence measure, several experiments were conducted on the static version of the model across varying numbers of observations used to construct the likelihood. \autoref{fig:Results/Static_comparation.png} shows the results where increasing the number of observations leads to improved performance. However, this improvement comes at the cost of increased computational effort, as greater time is required for inference. Given that the goal is to run inference overnight, these results can guide the choice of an appropriate \(N_{\text{max}}\) for the dynamic version by balancing feasible computation time with accuracy.

\textbf{Dynamic model.} In the following experiments, the number of observations used to construct the likelihood at each iteration is set to \(500\), \(1{,}000\), and \(2{,}000\). These values provide a balance between computational tractability for testing and the ability to draw meaningful conclusions about performance in larger, operational-scale settings.

\begin{figure}[tbp]
    \centering
    \includegraphics[width=0.6\linewidth]{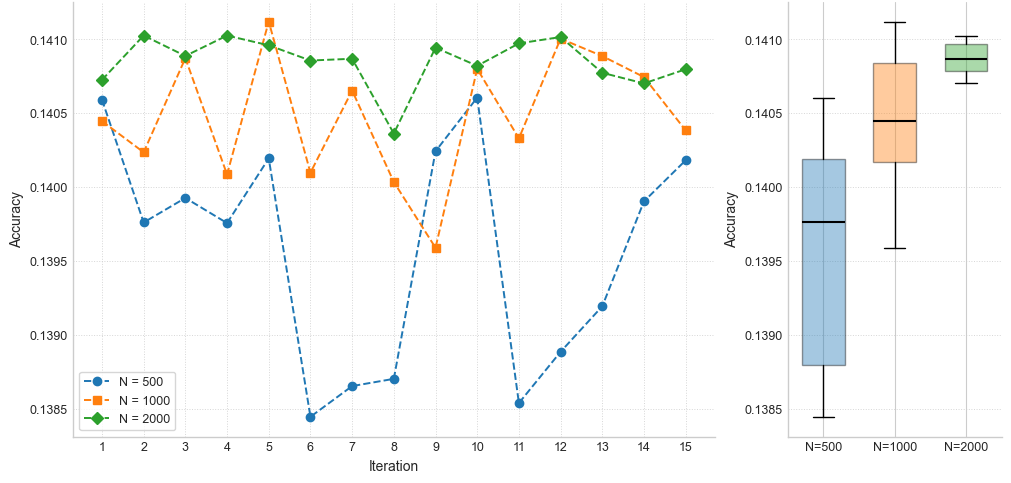}
    \caption[Filtering-step performance for the dynamic model]{Accuracy of filtering step in dynamic model over 15 iterations. Observation weights are set to $\beta=0$, corresponding to a no-memory setting: past observations are discarded after being used, and only current batch contributes to MCMC update. (left) Accuracy over iterations. (right) Accuracy boxplot for different number of observations across 15 iterations.}
    \label{fig:Results/Dynamic_Nobs_NotMean_notitle.png}
    \vspace{-0.2cm}
\end{figure}

\autoref{fig:Results/Dynamic_Nobs_NotMean_notitle.png} summarizes the behavior of the dynamic model over multiple iterations for different numbers of observations. In this experiment, corresponding to a no-memory weighting scheme, $\beta=0$, the weight assigned to each particle is $\eta=1$ for observations in the current iteration and $\eta=0$ for observations from previous iterations. As expected, using more observations yields better predictive performance. Moreover, the accuracy trajectories for larger number of observations exhibit smaller oscillation and a higher central tendency across iterations. This pattern is also evident in the boxplots in \autoref{fig:Results/Dynamic_Nobs_NotMean_notitle.png}: an increasing number of observations leads to a higher median accuracy and reduced variability across iterations, indicating more stable behaviour which can be extrapolated to higher number of observations.

\begin{figure}[h]
\centering

\begin{minipage}[t]{0.49\linewidth}
    \centering
    \includegraphics[width=\linewidth]{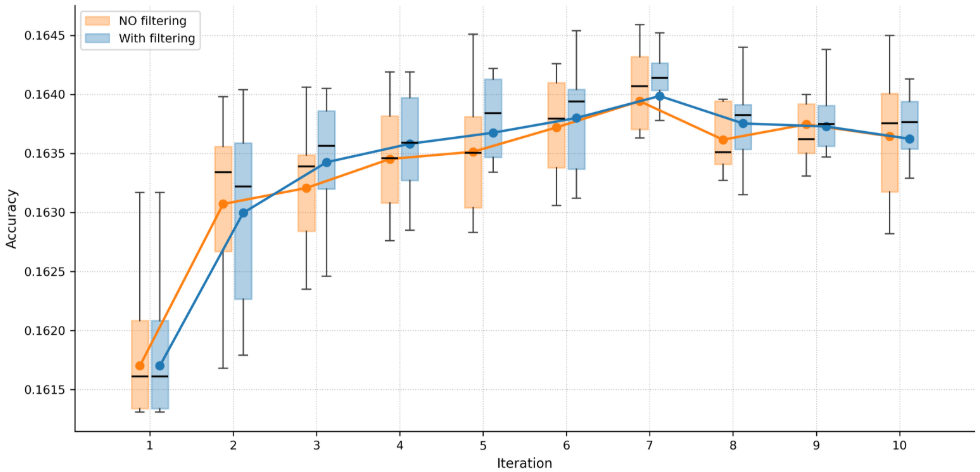}
    \captionof{figure}[Filtering-step performance for the dynamic model]{Comparison of accuracy of dynamic version over 10 iterations with (\textit{blue}) and without (\textit{orange}) filtering step. Eight runs were carried out to account for variability. Observation weights set to $\beta=0$, corresponding to a no-memory setting.}
    \label{fig:filtering_nolegend}
\end{minipage}
\hfill
\begin{minipage}[t]{0.49\linewidth}
    \centering
    \includegraphics[width=\linewidth]{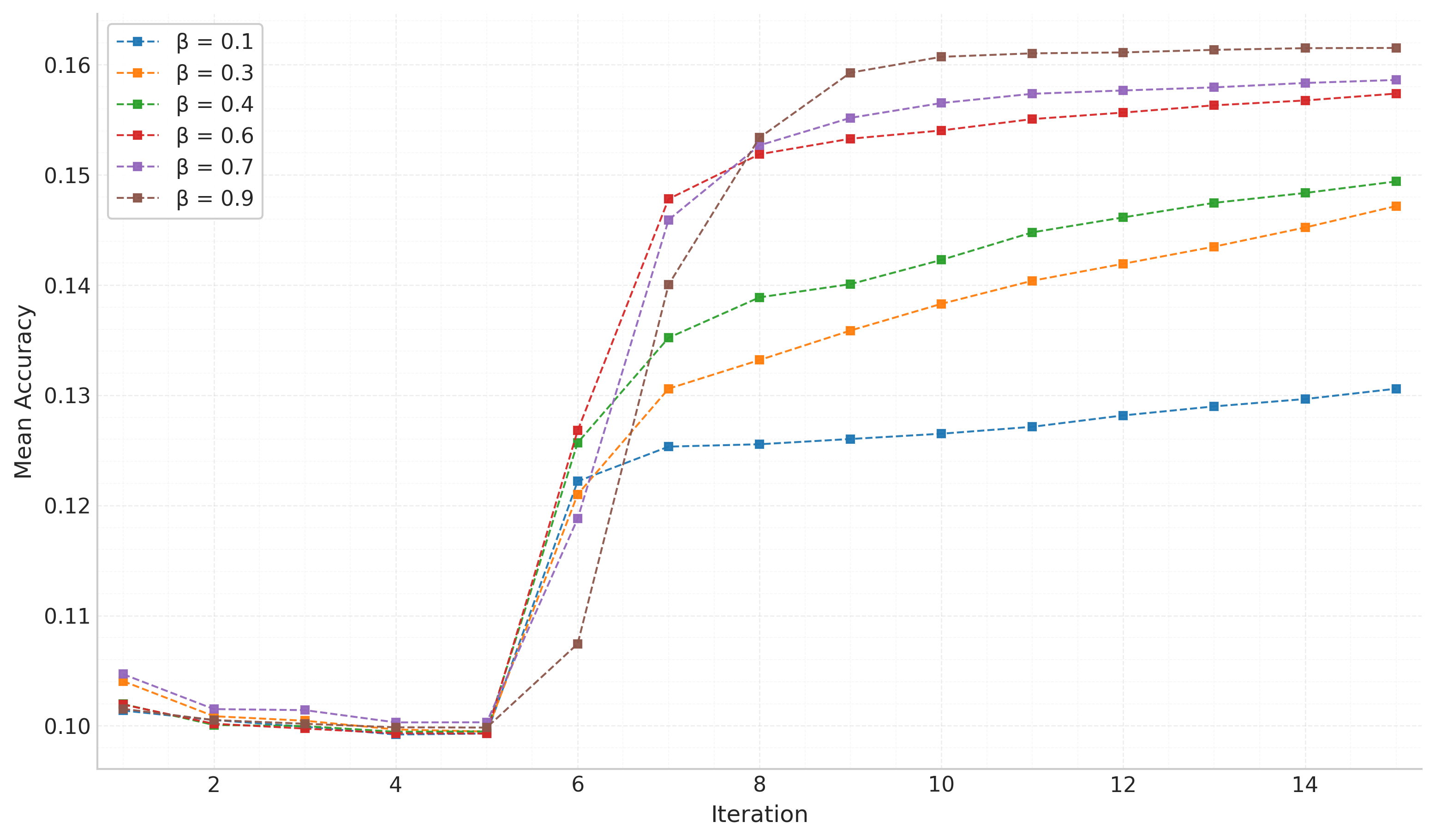}
    \captionof{figure}{Accuracy over 15 iterations for different $\beta$ values, with a forced change in misleading data after fourth iteration to test convergence.}
    \label{fig:betadependence}
\end{minipage}

\end{figure}

In addition, a separate simulation was conducted using likelihoods built from \(2{,}500\) observations, comparing ten runs with and without the filtering step over 10 iterations. As shown in \autoref{fig:filtering_nolegend}, both approaches start with similar accuracy, but the version with filtering performs slightly better in the long run, both in mean and variance. This is consistent with the role of filtering, which concentrates probability mass on more plausible particles, discards low-support ones, and thereby stabilizes inference and improves predictive accuracy over time. This effect is also visible in the shorter boxplots lower--tails across iterations.

Finally, we evaluate how convergence depends on the weighting scheme. The results reported here use weights that depend only on the age of the observations, \(\eta_i = \beta^{\alpha_i}\), where \(\alpha_i \in \{1,2,\ldots\}\) denotes the age of the \(i\)-th observation. The experiment begins with several iterations under misleading data and then imposes a shift at a fixed time point. At each iteration, \(2{,}500\) weighted observations are added to the likelihood, and a maximum age \(A_{\text{max}}=5\) is used. This setup allows us to assess how different values of \(\beta\) affect the model’s ability to forget outdated information and adapt to new conditions.

\autoref{fig:betadependence} shows that smaller values of \(\beta\) forget past information more quickly, improving accuracy immediately after the change, but reduce long-run performance. Larger values adapt more slowly, but typically achieve better asymptotic accuracy once the system stabilizes. In practice, these results suggest choosing \(\beta\) in the range \([0.9,1)\). A natural extension would be to make \(\beta\) adaptive, using higher values when data are scarce or the environment is stable, and lower values when a change is detected.

\FloatBarrier

\section{Use cases}

This section presents four relevant use cases for the proposed preference model within the smart mobility platform: (i) an automatic route-selection system that identifies the best routes to be displayed to a platform user; (ii) an automatic car-pooling recommendation system, which relies on a seat-offering mechanism aimed at both active and potential users; (iii) a mechanism to compute personalized and well-calibrated incentives; and (iv)  an auxiliary tool to generate synthetic data. In all four applications, decisions are driven by the acceptance probability
$p\!\left(\mathbf{w}(\mathbf{a}, \mathbf{z}^i), \mathbf{c}^i, \mathbf{t}^i, \mathbf{tw}^i\right)$,
or, equivalently, by the \textit{value}, $v(t, t_w, c) = - w_1 t - w_2 c - w_3 t_w$:
the greater the value of these quantities, the more likely the route is to be selected.

\FloatBarrier

\subsection{Relevant route offer selection}
\label{subsec: A predictive PL-based routine for choosing relevant routes}
The smart mobility platform includes a route generation system (OTP2) that combines public transport modes (bus, taxi, underground, train, etc.) with car-pooling services. For a user with features \(\mathbf{z}\), a query to the system returns \(n\) potential non-dominated routes in terms of cost and duration between the specified origin and destination within the required travel time window. From this set, a subset of \(K\) routes must be selected, ensuring that the displayed options are sufficiently differentiated to provide distinct alternatives, where \(K\) is fixed by system designers.

To address this problem, the following procedure is applied:
\begin{itemize}
\item The set of routes is partitioned into \(K\) clusters.
\item Within each cluster, the route most likely to be preferred by the user is identified.
\end{itemize}

\begin{figure}[ht]
\centering
\begin{tikzpicture}[
  font=\small,
  box/.style={
    draw,
    rounded corners=6pt,
    line width=0.7pt,
    inner sep=2pt,
    outer sep=0pt,
    align=center
  },
  caption/.style={font=\small\bfseries, align=center}
]

\def\gap{0.8cm}  

\node[box] (b1) {\includegraphics[width=4.2cm,height=2.7cm,keepaspectratio]{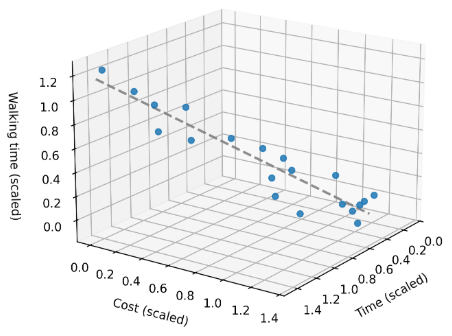}};
\node[box, right=\gap of b1] (b2) {\includegraphics[width=4.2cm,height=2.7cm,keepaspectratio]{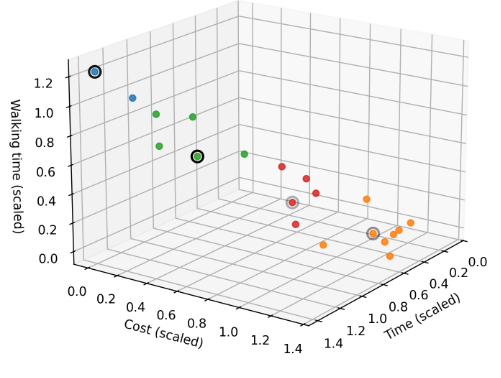}};
\node[box, right=\gap of b2] (b3) {\includegraphics[width=4.2cm,height=2.2cm,keepaspectratio]{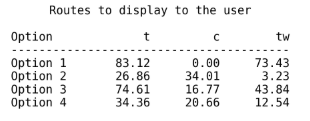}};

\node[caption, above=0.5pt of b1] {Input: $\mathbf{n}$ Possible routes};
\node[caption, above=0.5pt of b2] {Clustering};
\node[caption, above=0.5pt of b3] {$\mathbf{K=4}$ prefered selected routes};

\def\ah{0.26cm}   
\def\at{0.22cm}   
\def\pad{0.06cm}  

\coordinate (A1) at ($(b1.east)+(\pad,0)$);
\coordinate (B1) at ($(b2.west)-(\pad,0)$);
\filldraw[draw=black, fill=gray!15, line width=0.9pt]
  ($(A1)+(0,\ah)$) -- ($(B1)+(-\at,\ah)$) -- ($(B1)+(-\at,1.6*\ah)$)
  -- (B1) -- ($(B1)+(-\at,-1.6*\ah)$) -- ($(B1)+(-\at,-\ah)$)
  -- ($(A1)+(0,-\ah)$) -- cycle;

\coordinate (A2) at ($(b2.east)+(\pad,0)$);
\coordinate (B2) at ($(b3.west)-(\pad,0)$);
\filldraw[draw=black, fill=gray!15, line width=0.9pt]
  ($(A2)+(0,\ah)$) -- ($(B2)+(-\at,\ah)$) -- ($(B2)+(-\at,1.6*\ah)$)
  -- (B2) -- ($(B2)+(-\at,-1.6*\ah)$) -- ($(B2)+(-\at,-\ah)$)
  -- ($(A2)+(0,-\ah)$) -- cycle;

\end{tikzpicture}
\caption{Overview of route selection process. First, $n=20$ routes are generated by the smart mobility system. Second, the routes are clustered into $K=4$ groups. Finally, the routes predicted to be most preferred at each cluster are displayed to the user.}
\label{fig:pipeline-tikz}
\end{figure}
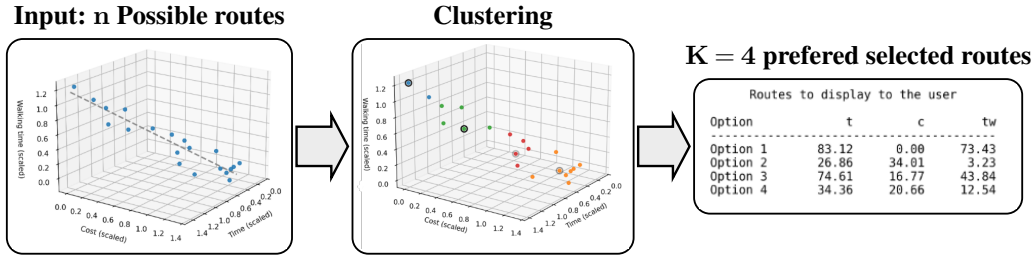
\begin{algorithm}[ht]
\caption{\textbf{Route Selection}}
\label{alg:clustered-route-selection}

\begin{algorithmic}[1]
\vspace{-0.2cm}
\Statex \hspace{-0.75cm}\Statex \hspace{-0.75cm}\textbf{Input:} number $\mathbf{K}$ of routes to be displayed, $n$ potential routes described by $\mathbf{\{(t_i,c_i,t_{w_i})\}_{i=1}^n}$, customer features vector $\mathbf{z}$ and parameter $\mathbf{\hat{a}}$ posterior means.

\Statex \hspace{-0.75cm} \textbf{Output:} $\mathbf{K}$ selected routes.
\vspace{-0.2cm}
\Statex \hspace{-0.75cm} \rule{1.05\linewidth}{0.4pt}

\State \textit{\textbf{Initialisation:}}
\Statex \quad 1.1: Compute weight vector ${\bf w}(\hat{a} z) = (w_1,w_2,w_3)$

\State \textit{\textbf{Clustering} (\texttt{k-means}):}
\Statex \quad 2.1: Cluster the $n$ routes into $K$ groups based on  features  $(t_i,c_i,t_{w_i})$.

\State \textit{\textbf{Route Selection:}}
\Statex \quad 3.1: \textbf{for} $i = 1, \dots, K$ \textbf{do}
\Statex \quad \quad\quad Within the $i$-th cluster, compute utility for every proposed route,
\Statex \quad \quad\quad \quad \quad \quad $v = -w_1(\hat{a}z) \cdot t - w_2(\hat{a}z) \cdot c - w_3(\hat{a}z) \cdot t_w$
\Statex \quad \quad\quad Select the route with highest utility.
\Statex \quad 3.2: end \textbf{for}

\end{algorithmic}
\end{algorithm}

\noindent Since the system must return the \(K\) alternatives in real time, fully integrating over the posterior distribution to evaluate the predictive probability of each route being preferred is computationally infeasible. Instead, this probability is approximated by evaluating it at the posterior mean \(\bar{a}\), and selecting the route with the highest predicted choice probability within each cluster. Repeating this procedure across all clusters provides a set of diverse alternatives, as described in \autoref{alg:clustered-route-selection} and illustrated in \autoref{fig:pipeline-tikz}.

\FloatBarrier

\subsection{Proactive coordinated car pooling services}

The proposed design is \emph{model-driven}: the trained preference model estimates the value of the car-pooling route for each potential user, and the system then prioritizes users with higher estimated value, that is, those who are more likely to accept the car-pooling option. Rides are filled quickly via a waiting list and time-limited offers that are reissued upon rejection or non-response, while maintaining vetoed users excluded.\footnote{The system includes a veto capability to indicate that a user does not wish to share rides with a specific driver or rider in later routes.} When demand is insufficient to fill a trip, incentives might be considered so as to encourage potential users.

\begin{algorithm}[ht]
\caption{\textbf{Ride Assignment in Coordinated car-pooling}}
\label{alg:ride-assignment}
\small
\begin{algorithmic}[1]

\State \textbf{\textit{Ride registration:}} Register ride attributes (schedule/location) and public-transport  connections.
\State \textbf{\textit{Customer filtering:}} Build \texttt{Customer\_List} of users satisfying temporal/spatial proximity and mutual non-veto constraintss.

\State \textbf{While} \texttt{Car\_Capacity} $> 0$ \textbf{do}
\Statex \quad \textbf{If} \texttt{Customer\_List(reduced mobility)} is not empty \textbf{then}
\Statex \qquad \textbf{\textit{Priority 1: reduced mobility:}} rank by \textit{increasing walking time} and offer rides until list empty.
\Statex \quad \textbf{Else if} \texttt{Customer\_List(non-reduced mobility)} is not empty \textbf{then}
\Statex \qquad \textbf{\textit{Priority 2: emissions savings:}} rank by emission savings and offer rides until full or list empty.
\Statex \quad \textbf{Else}
\Statex \qquad \textbf{\textit{Priority 3: Preference model:}} use the trained model to compute the value function,
\[
    \texttt{value} \;=\; -w_1(\boldsymbol{\hat{a}z})\,t
    \;-\; w_2(\boldsymbol{\hat{a}z})\,c
    \;-\; w_3(\boldsymbol{\hat{a}z})\,t_w .
\]
\Statex \qquad Rank by increasing \texttt{value}.
\Statex \qquad \textbf{If} \texttt{Customer\_List} is empty \textbf{then break}
\Statex \quad \textbf{End if}
\State \textbf{End while}

\State \textbf{If} a rider cancels before departure \textbf{then} penalize rider \texttt{value}, reintroduce car, increment \texttt{Car\_Capacity}, and restart assignment.

\end{algorithmic}
\end{algorithm}

\begin{figure}[ht]
\centering
\begin{tikzpicture}[
  font=\normalsize, 
  box/.style={
    draw,
    rounded corners=7pt,
    line width=0.85pt,
    inner sep=3pt,     
    outer sep=0pt,
    align=center
  },
  caption/.style={
    font=\normalsize\bfseries,
    align=center
  }
]

\def\gap{1.2cm}   

\node[box] (b1) {\includegraphics[width=4.2cm,height=2.7cm,keepaspectratio]{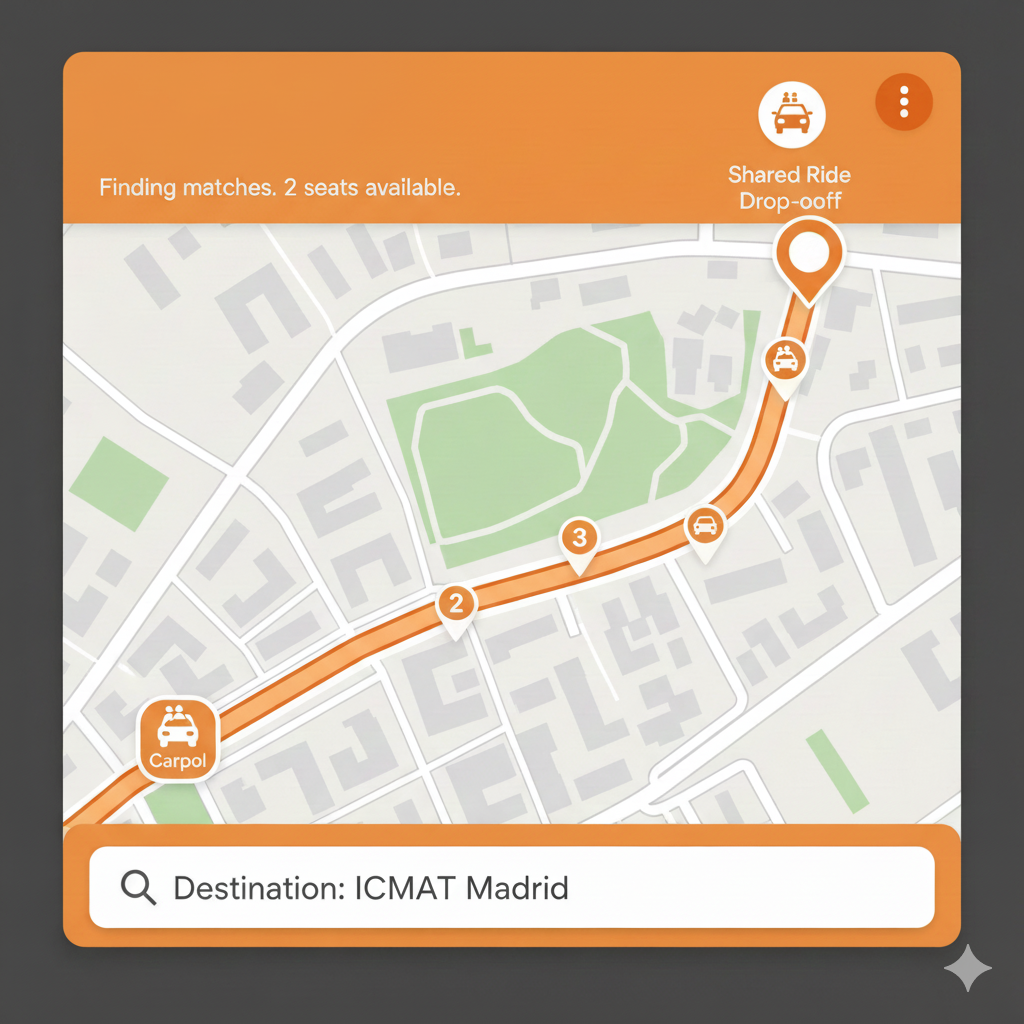}};
\node[box, right=\gap of b1] (b2) {\includegraphics[width=4.2cm,height=2.7cm,keepaspectratio]{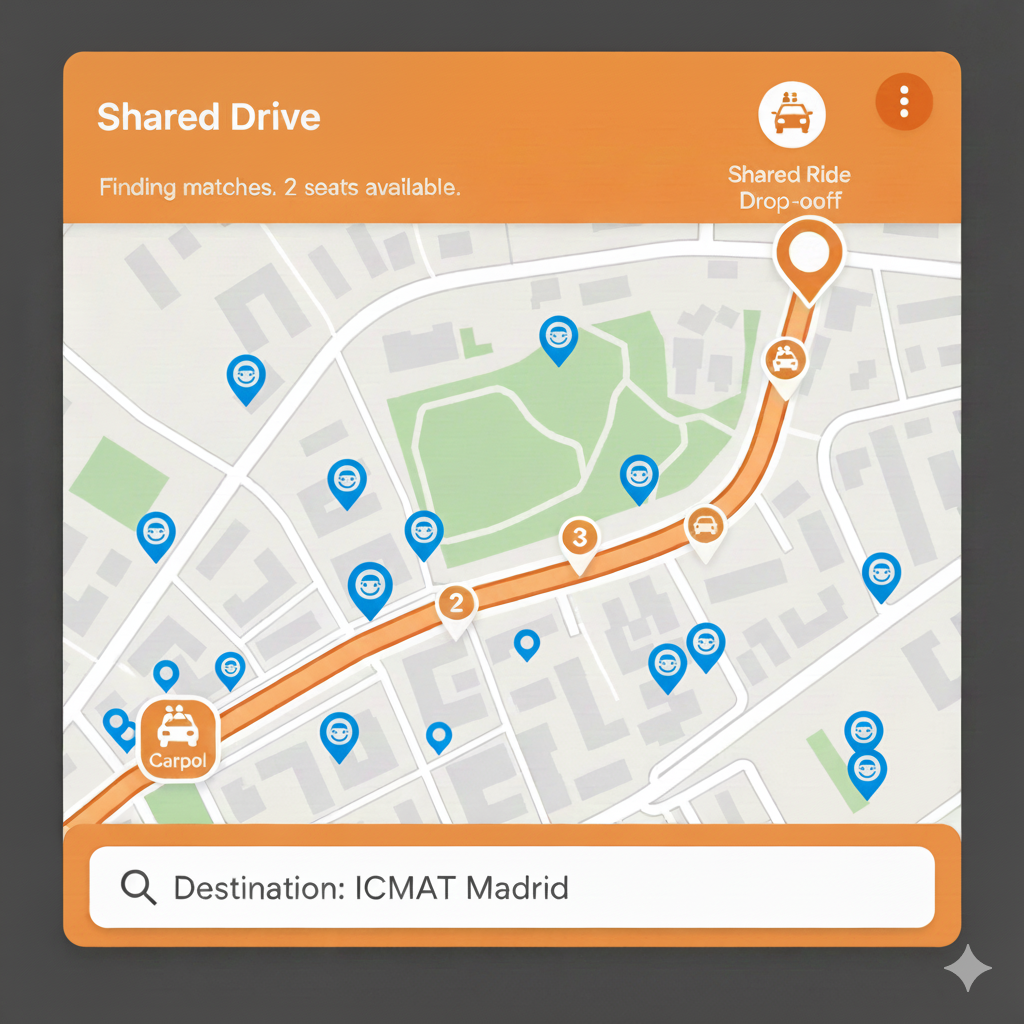}};
\node[box, right=\gap of b2] (b3) {\includegraphics[width=4.2cm,height=2.2cm,keepaspectratio]{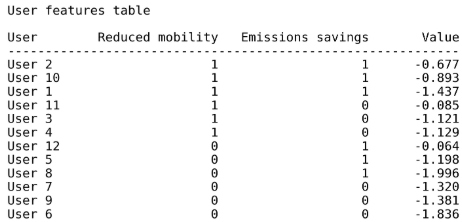}};

\node[caption, above=1pt of b1] {Input: Car ride};
\node[caption, above=1pt of b2] {Identify potential users};
\node[caption, above=1pt of b3] {Rank users};

\def\ah{0.30cm}   
\def\at{0.26cm}
\def\pad{0.17cm}

\coordinate (A1) at ($(b1.east)+(\pad,0)$);
\coordinate (B1) at ($(b2.west)-(\pad,0)$);
\filldraw[draw=black, fill=gray!15, line width=1.0pt]
  ($(A1)+(0,\ah)$) -- ($(B1)+(-\at,\ah)$) -- ($(B1)+(-\at,1.6*\ah)$)
  -- (B1) -- ($(B1)+(-\at,-1.6*\ah)$) -- ($(B1)+(-\at,-\ah)$)
  -- ($(A1)+(0,-\ah)$) -- cycle;

\coordinate (A2) at ($(b2.east)+(\pad,0)$);
\coordinate (B2) at ($(b3.west)-(\pad,0)$);
\filldraw[draw=black, fill=gray!15, line width=1.0pt]
  ($(A2)+(0,\ah)$) -- ($(B2)+(-\at,\ah)$) -- ($(B2)+(-\at,1.6*\ah)$)
  -- (B2) -- ($(B2)+(-\at,-1.6*\ah)$) -- ($(B2)+(-\at,-\ah)$)
  -- ($(A2)+(0,-\ah)$) -- cycle;

\end{tikzpicture}
\caption{Overview of proactive coordinated car-pooling process. A new ride is recorded. Next, potential customers are identified. Finally, they are ranked based on reduced mobility, emission reduction, and preference-model value.}
\label{fig:carsharing}
\vspace{-0.2cm}
\end{figure}
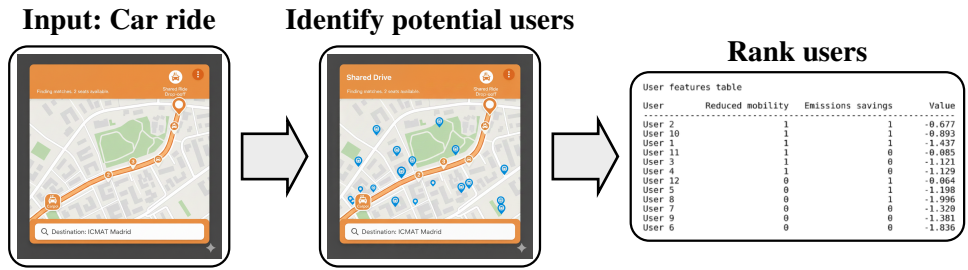

\noindent The proposed approach, illustrated in \autoref{alg:ride-assignment} and schematized in \autoref{fig:carsharing}, assigns seats in a coordinated car-pooling environment based on a priority-based ranking. When a vehicle becomes available, the platform registers the trip attributes (route, departure time, location, and capacity) and stores contextual information such as nearby public-transport  connections that can be used to estimate alternative travel options.

The system then constructs the compatible candidate set by filtering users who satisfy temporal and spatial proximity constraints,
   also removing any disallowed matches due to veto rules. The resulting \texttt{Customer\_List} contains only users who can be offered a seat without violating the platform or user constraints.

Seat allocation proceeds iteratively while capacity remains open, following a three-stage priority logic. First, reduced-mobility riders are prioritized and ranked by increasing walking time to minimize access burden. If no such riders remain, candidates are ranked by decreasing expected emissions savings. Finally, the remaining candidates are ordered to maximize acceptance using the learned preference model: for each user $i$ the platform evaluates $v(t,t_w,c) = - w_1(\boldsymbol{\hat{a}z})\,t - w_2(\boldsymbol{\hat{a}z})\,c - w_3(\boldsymbol{\hat{a}z})\,t_w,$
and prioritizes users with higher value.

\FloatBarrier

\subsection{Incentive design}
Improving a transport  system often requires encouraging certain users to adjust their behavior to benefit the overall network. For instance, incentives may be used to reduce emissions, promote the use of public transport, or ensure better service for passengers with reduced mobility. Consider a driver who follows a preferred route, while a passenger with reduced mobility would benefit from being picked up at a different location. The difficulty is that accommodating this passenger requires the driver to deviate from its preferred route, increasing travel time by an amount the driver is not willing to accept. In such cases, the platform can offer a monetary incentive that shifts the driver’s decision. To design this mechanism, we rely on our preference model: the compensation offered should adapt to the specific driver we aim to influence.

Suppose that there are two candidate routes, $l_1$ and $l_2$. Route $l_1$ corresponds to the driver’s baseline trip, while route $l_2$ involves a detour that increases travel time in order to serve the passenger with reduced mobility. Following \autoref{eq:pl_softmax2}, the odds of preferring $l_1$ over $l_2$ can be written as
\begin{equation}
\text{odds}\!\left(l_1 \succ l_2\right)
=
\frac{\exp\!\left(v(\mathbf{r}_{l_1})\right)}{\exp\!\left(v(\mathbf{r}_{l_2})\right)}
=
\exp\!\Big(-w_t\,(t_{l_1}-t_{l_2}) - w_c\big(c_{l_1}-(c_{l_2}-I)\big)\Big),
\label{eq:pl_incentive}
\end{equation}
where $I$ denotes the incentive offered to the driver. To ensure that the detour route is at least as attractive as the baseline, we require $\text{odds}\!\left(l_1 \succ l_2\right)=1$, yielding the minimum incentive
\begin{equation*}
I = \frac{w_t}{w_c}\,(t_{l_2}-t_{l_1}) - (c_{l_1}-c_{l_2}).
\end{equation*}
This expression highlights how the learned weights, which quantify the relative importance a given driver assigns to time and cost, directly translate into a personalized incentive. In particular, drivers who strongly prioritize time over cost ($w_t \gg w_c$) require larger incentives since $\tfrac{w_t}{w_c} \gg 1$. Similarly, larger time differences between both routes, $(t_{l_2}-t_{l_1})$, imply higher compensation. The same personalized incentive mechanism can be applied to other objectives, such as discourage high-emission routes or encourage users to switch to public transport  to avoid traffic congestion in certain areas. 

More broadly, this idea can be embedded in an optimization problem that accounts for multiple users simultaneously, e.g., to steer drivers towards routes that are most socially valuable. Let $v_j(\mathbf{r}_{l_i})$ denote the utility that user $j$ associates with choosing the ride-pooling option $l_i$. Here, each $l_i$ represents a candidate route presented to the driver, and different routes may benefit different subsets of users. We then define the expected total utility by offering incentive $I$ as
\begin{equation*}
\mathbb{E}_I[U]
=
\sum_i 
\mathbb{P}\!\left(d \text{ accepts } l_i \mid I\right)
\Bigg(
\prod_{j \in \mathcal{J}(l_i)} \mathbb{P}\!\left(j \text{ accepts } l_i\right)
\Bigg)
\Bigg(\sum_{j \in \mathcal{J}(l_i)} v_j(\mathbf{r}_{l_i})+v_d(\mathbf{r}_{l_i})\Bigg),
\end{equation*}
where $\mathcal{J}(l_i)$ denotes the set of users served by route $l_i$ and $d$ is the driver. The driver acceptance probability can be modeled using \autoref{eq:pl_incentive}, while the remaining terms capture the acceptance behavior of prospective passengers.
By maximizing $\mathbb{E}_I[U]$, we obtain the incentive that yields the highest expected social benefit. In this way, the framework does not only account for the incentive required to shift the driver’s decision, but also for how that incentive contributes to maximizing the overall social utility.

\FloatBarrier

\subsection{Synthetic data generation}
\label{sec:Data Generation Process}

The goal is to simulate user behavior and route preferences in a manner that reflects plausible real-world patterns. This would allow developers to understand the underlying data structures and flows to apply them in future applications.

The desired dataset consists of $n$ daily user interactions. Each of them includes the observed data marked in grey in \autoref{fig:Modeldiagram}: $K$ different routes, characterized by their time, cost, and time walked; the features $\mathbf{z}$ of the user and its context; and the user’s choice $i_0$, indicating their preferred route. Several considerations are taken into account to ensure that the data generation process yields realistic user interactions. 

To generate a single observation, it is necessary to generate $k$ route options with their corresponding attributes $\{(t_i, c_i, t_{w_i})\}_{i=1}^k$. These features are inherently correlated. For instance, a route with a shorter total time may incur higher costs due to the use of faster transport  modes, such as taxis or premium services. Additionally, the walking time must not exceed the total transport  time, $t_{w_i} \leq t_i$ for all $i$. Therefore to simulate each iteration, the following steps are taken.

\begin{algorithm}[h]
\caption{\textbf{Synthetic Route-Choice Generation}}
\label{alg:synthetic-route-choice}
\small
\begin{algorithmic}[1]
\Statex \hspace{-0.5cm}\textbf{Input:} Parameter matrix $\mathbf{a}\in\mathbb{R}^{2\times 7}$.
\Statex \hspace{-0.65cm} \textbf{Output:} One user interaction:$\big\{\{(t_i,t_{w_i},c_i)\}_{i=1}^k, \mathbf{z}=(z_{11},z_{12},z_{13},z_2,z_3,z_4,z_5),i_0\big\}$.
\Statex \hspace{-0.75cm } \rule{1.05\linewidth}{0.4pt}  

\State \textbf{\textit{Draw upper bound constraints:}}
\[
\texttt{max}_{\texttt{time}}\sim \mathcal{U}(10,80)\qquad
\texttt{max}_{\texttt{walkedtime}}\sim \mathcal{U}(0,\texttt{max}_{\texttt{time}})\qquad
\texttt{max}_{\texttt{cost}}\sim \mathcal{U}(10,40)
\]

\State \textbf{\textit{Route attribute generation:}}
\Statex \textbf{For} each route $i\in\{1,\dots,K\}$ \textbf{do}
\Statex \quad Sample $u_i \sim \mathcal{U}(0,1)$.
\[
t_i \;=\; u_i\,\texttt{max}_{\texttt{time}} + \varepsilon^t_i,\quad
t_{w_i} \;=\; u_i\,\texttt{max}_{\texttt{walkedtime}} + \varepsilon^{tw}_i,\quad
c_i \;=\; (1-u_i)\,\texttt{max}_{\texttt{cost}}+ \varepsilon^c_i,
\]
\Statex \quad  where $\varepsilon^t_i,\varepsilon^{tw}_i,\varepsilon^c_i \sim \mathcal{N}(0,\sigma^2)$
with $\sigma = 0.002\cdot \texttt{maxtime}\cdot \texttt{maxcost}$,
\Statex \textbf{End for}

\State \textbf{\textit{Remove strictly dominated routes:}} Remove any route $i$ for which there exists another route $j$ such that $c_j < c_i$ \textbf{and} $t_j < t_i$. If removed, generate new $(t_i,c_i,{t_w}_i)$

\State \textbf{\textit{Generate user/context features independently:}} \Statex \quad $z_1=(z_{11},z_{12},z_{13})$ as one-hot with $\mathbb{P}(z_{1j}=1)=\tfrac{1}{3}$. \Statex \quad $z_2 \sim \mathcal{U}(0,1)$. \Statex \quad $z_3 \in \{0,1\}$ with $\mathbb{P}(z_3=1)=\tfrac{1}{2}$. \Statex \quad $z_4 \sim \mathcal{U}(0,1)$. \Statex \quad $z_5 \in \{0,1,2\}$ with $\mathbb{P}(z_5=m)=\tfrac{1}{3}$ for $m\in\{0,1,2\}$.

\State \textbf{\textit{Sample choice:}} Randomly sample the chosen route from the resulting probabilities distribution over the  routes.
\end{algorithmic}
\end{algorithm}

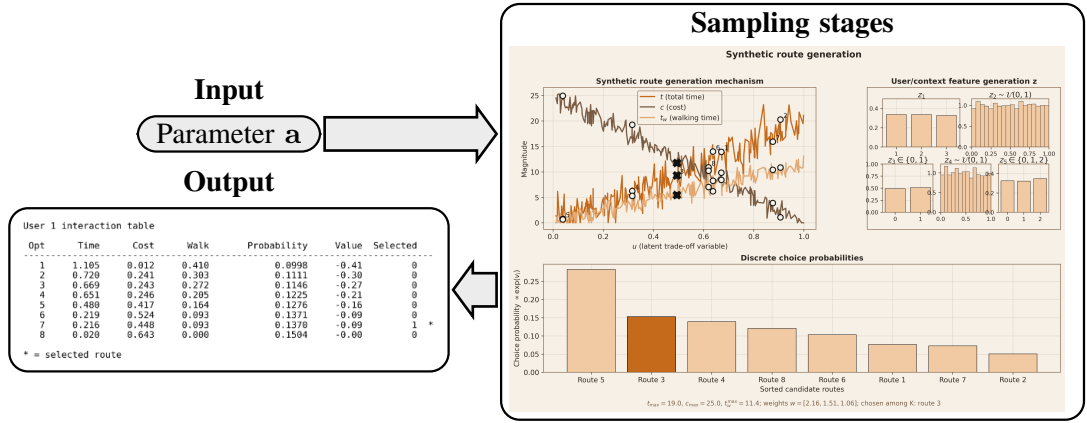
\begin{figure}[ht]
\centering
\begin{tikzpicture}[
  font=\normalsize, 
  box/.style={
    draw, rounded corners=7pt,
    line width=0.85pt,
    inner sep=3pt,  
    outer sep=0pt,
    align=center
  },
  caption/.style={font=\normalsize\bfseries, align=center}
]

\node[
  box, fill=gray!15,
  text width=2.2cm, align=center
] (b1) {Parameter $\mathbf{a}$};
\node[caption, above=1pt of b1] {Input};

\node[
  box,
  below=0.75cm of b1
] (b3) {\includegraphics[width=5.6cm,height=3.2cm,keepaspectratio]{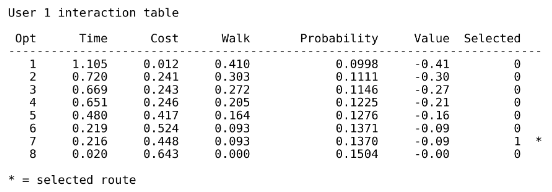}};
\node[caption, above=1pt of b3] {Output};

\path (b1.center) -- (b3.center) coordinate[midway] (midL);

\node[
  box,
  right=3.6cm of midL, 
  anchor=west,
  text width=7.5cm,
  minimum height=5.2cm, 
  align=left
] (b2) {%
\centering
\textbf{Sampling stages}\\[2pt]
\includegraphics[width=0.98\linewidth]{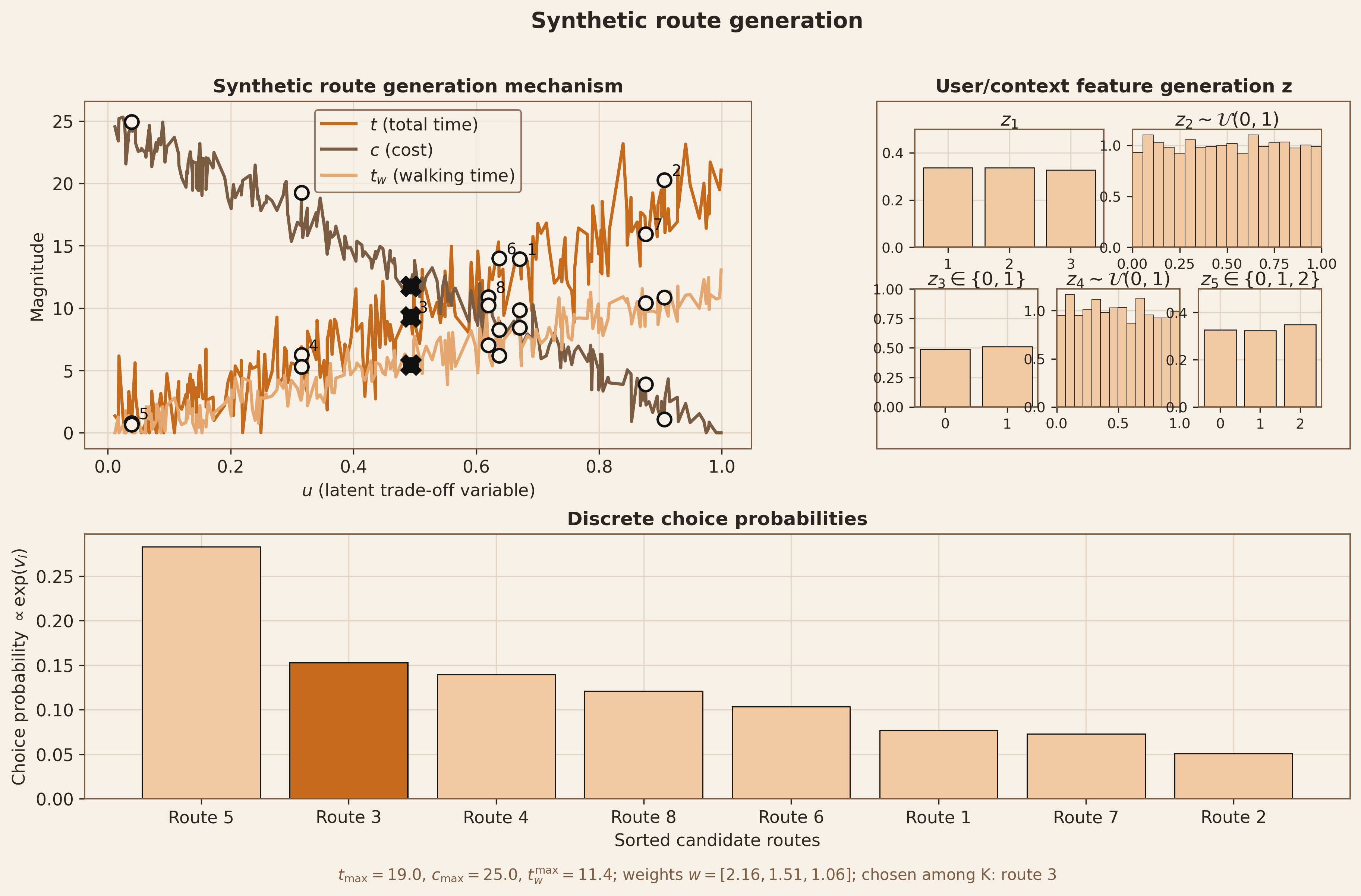}
};

\def\ah{0.22cm}  
\def\at{0.26cm}
\def\pad{0.06cm}

\coordinate (A1) at ($(b1.east)+(\pad,0)$);
\coordinate (B1) at ($(b2.west |- b1.center)-(\pad,0)$);
\filldraw[draw=black, fill=gray!15, line width=1.0pt]
  ($(A1)+(0,\ah)$) -- ($(B1)+(-\at,\ah)$) -- ($(B1)+(-\at,1.6*\ah)$)
  -- (B1) -- ($(B1)+(-\at,-1.6*\ah)$) -- ($(B1)+(-\at,-\ah)$)
  -- ($(A1)+(0,-\ah)$) -- cycle;

\coordinate (A2) at ($(b2.west |- b3.center)-(\pad,0)$);
\coordinate (B2) at ($(b3.east)+(\pad,0)$);
\filldraw[draw=black, fill=gray!15, line width=1.0pt]
  ($(A2)+(0,\ah)$) -- ($(B2)+(\at,\ah)$) -- ($(B2)+(\at,1.6*\ah)$)
  -- (B2) -- ($(B2)+(\at,-1.6*\ah)$) -- ($(B2)+(\at,-\ah)$)
  -- ($(A2)+(0,-\ah)$) -- cycle;

\end{tikzpicture}
\caption{Synthetic data generation process.}
\label{fig:datagen}
\end{figure}

\noindent First, the algorithm draws upper bounds to set the scale of the interaction: a maximum travel time ($\texttt{max}_{\texttt{time}}$) is sampled uniformly from $[10,80]$, a maximum walked time ($\texttt{max}_{\texttt{walkedtime}}$) is sampled uniformly from $[0,\texttt{max}_{\texttt{time}}]$, and a maximum cost ($\texttt{max}_{\texttt{cost}}$) is sampled uniformly from $[10,40]$.\footnote{These range of values were consulted with transport  experts. However, they can be adjusted to match a desired transport  environment.} The constraint $\texttt{max}_{\texttt{walkedtime}}\leq \texttt{max}_{\texttt{time}}$ guarantees that walked time remains compatible with total travel time. Given these bounds, the algorithm generates $K$ candidate routes by sampling a latent factor $u_i\sim\mathcal{U}(0,1)$ for each route, using it to jointly construct the attributes
\[
t_i=u_i\,\texttt{max}_{\texttt{time}}+\varepsilon_i^t,\qquad
t_{w_i}=u_i\,\texttt{max}_{\texttt{walkedtime}}+\varepsilon_i^{tw},\qquad
c_i=(1-u_i)\,\texttt{max}_{\texttt{cost}}+\varepsilon_i^c,
\]
with perturbations $\varepsilon_i^t,\varepsilon_i^{tw},\varepsilon_i^c\sim\mathcal{N}(0,\sigma^2)$ and $\sigma=0.002\cdot\texttt{max}_{\texttt{time}}\cdot\texttt{max}_{\texttt{cost}}$.\footnote{The noise level is chosen to scale with the magnitude of the route attributes: larger times and costs imply higher variance, while the factor $0.002$ ensures that the perturbations remain at the same order of magnitude as the corresponding attribute values.} Reusing the same $u_i$ across attributes induces correlation while the additive Gaussian noise avoids deterministic relationships. After generating the $K$ routes, strictly dominated alternatives are removed, as well as any routes with negative travel times or costs.\footnote{A route $i$ is considered dominated if there exists another route $j$ such that $c_j<c_i$ and $t_j<t_i$.} Moreover, whenever a route is removed, new attributes $(t_i,t_{w_i},c_i)$ are regenerated until no strictly dominated routes remain. In parallel, user and contextual features are sampled independently to form $\mathbf{z}=(z_{11},z_{12},z_{13},z_2,z_3,z_4,z_5)$: $z_1=(z_{11},z_{12},z_{13})$ is a three-class one-hot vector with $\mathbb{P}(z_{1j}=1)=1/3$, $z_2\sim\mathcal{U}(0,1)$, $z_3\in\{0,1\}$ with $\mathbb{P}(z_3=1)=1/2$, $z_4\sim\mathcal{U}(0,1)$, and $z_5\in\{0,1,2\}$ with $\mathbb{P}(z_5=m)=1/3$ for $m\in\{0,1,2\}$.\footnote{If a specific application requires emphasizing particular classes or user types, these probabilities can be modified accordingly.}

Finally, given the resulting alternatives and context vector, the algorithm computes route-choice probabilities using the PL adaptation associated with~\autoref{fig:Modeldiagram} and the input parameters  $\mathbf{a}$. The chosen route index $i_0$ is then sampled from the resulting probability distribution to reflect realistic variability in user decisions. This stochastic selection better reflects realistic decision-making, where choices may be influenced by unobserved preferences or situational factors. As a result, the simulated dataset is more realistic and inherently noisy. For further details see \autoref{alg:synthetic-route-choice} and \autoref{fig:datagen}.

\FloatBarrier

\section{Discussion and Future Work}

This work has developed and assessed a model to facilitate user’s route choice prediction using personality features and contextual factors, with the aim of integrating it within an AI-based smart mobility platform. Applications include enhancing car-pooling services and improving route allocation systems. The proposed model extends Plackett–Luce framework to handle discrete choice data, and its Bayesian approach supports dynamic adaptation while maintaining reasonable computational efficiency.

The static model achieved acceptable performance, particularly when fitted using a large number of observations. However, it lacks adaptability and becomes computationally inefficient as it requires to be re-estimated daily using both historical and newly observed data. This limitation motivated the development of dynamic approaches. Among the proposed dynamic variants, incorporating a filtering step leads to more robust and stable behavior. To further capture time-dependent effects, observation weights were introduced. Overall, the proposed dynamic scheme strikes a balance between computational efficiency, adaptability to temporal evolution through sequential updating and weighting, and stability through filtering.

Several directions for future work remain open. For instance, additional explanatory variables may be missing, leaving a substantial amount of variability to unobserved factors. Another promising avenue is to refine the weighting scheme to capture seasonal effects. Moreover, the static model can serve as a practical fallback when sufficient historical data are available. In particular, if the dynamic procedure diverges due to atypical or misleading inputs, the static model may be used as a reset mechanism to restore a reliable baseline.

All simulations in this study were executed with a 12th Gen Intel(R) Core(TM) i7-1255U processor (CPU), operating at 1.70 GHz. In the production setting, access to a significantly more powerful computing infrastructure is available and, moreover, inference can be run overnight so that more computation time is available until responses are required. Consequently, processes tested here with around 2,500 observations can be enlarged to a much larger scale. The patterns observed in terms of error and fluctuation behavior should therefore be extrapolated, but their magnitude would be notably reduced. This increase in computational capacity is anticipated to lead to a general improvement in accuracy beyond the results presented in this analysis.

\FloatBarrier

\begin{acks}[Acknowledgments]
This paper is part of a broader initiative, \textit{Dynamo}, co-financed by the Centre for the Development of Industrial Technology (CDTI) and the European Regional Development Fund (ERDF). We also thank the Zerel team for their valuable ideas, recommendations, and discussions related to the topics addressed in this paper. M. Santos-Pascual acknowledges the Spanish Ministry of Science, Innovation and Universities for the FPU24/04169 Ph.D. scholarship.

Code and data are available in \url{https://github.com/MiguelSantPasc/TwoLevel_PlacketLucce.git} where synthetic data is used for privacy reasons.
\end{acks}
\bibliographystyle{imsart-nameyear} 
\bibliography{references}       


\end{document}